\documentclass{aa}
\usepackage{graphicx}
\def\simless{\mathbin{\lower 1pt\hbox
   {$\spose{\raise 5pt\hbox{$\char'074$}}\char'430$}}}
\def\simgreat{\mathbin{\lower 1pt\hbox
   {$\spose{\raise 5pt\hbox{$\char'076$}}\char'430$}}}
\def\simgreat{\gapp}
\def\simless{\lapp}
\def\lapp{\mathbin{\raise2pt \hbox{$<$} \hskip-9pt \lower4pt \hbox{$\sim$}}}
\def\gapp{\mathbin{\raise2pt \hbox{$>$} \hskip-9pt \lower4pt \hbox{$\sim$}}}

\begin{document}

\title{Nonradial and nonpolytropic astrophysical outflows\protect\\
         VIII. A GRMHD generalization for relativistic jets}

\titlerunning{Nonradial and nonpolytropic astrophysical outflows VIII.}

  \author{Z. Meliani
           \inst{1,2}
   \and  C. Sauty
           \inst{1}
   \and  N. Vlahakis
           \inst{3}
   \and  K. Tsinganos
           \inst{3}
    \and  E. Trussoni
           \inst{4}
          }

 \offprints{Z. Meliani \\ (zakaria.meliani@obspm.fr)\\
{\it Present address:} Max Planck Institute for Astrophysics, Box 1317, D-85741,
        Garching, Germany}
   \institute
        { Observatoire de Paris, LUTh, F-92190 Meudon, France
    \and  Universit\'e de Paris 7, APC, 2 place Jussieu, 75005 Paris, France
    \and  Section of Astrophysics, Astronomy \& Mechanics, Department
         of Physics and IASA, University of Athens,
          Panepistimiopolis GR-157 84 Zografos, Athens, Greece
    \and Istituto Nazionale di Astrofisica (INAF) - Osservatorio Astronomico
       di Torino, Strada Osservatorio 20, I-10025 Pino Torinese (TO), Italy}
   \date{Received ... / accepted ...}

   \abstract{
Steady axisymmetric outflows originating at the hot coronal
magnetosphere of a Schwarzschild black hole and surrounding
accretion disk are studied in the framework of general
relativistic magnetohydrodynamics (GRMHD). The assumption of
meridional self-similarity is adopted for the construction of
semi-analytical solutions of the GRMHD equations describing
outflows close to the polar axis. In addition, it is assumed that
relativistic effects related to the rotation of the black hole and
the plasma are negligible compared to the gravitational and other
energetic terms. The constructed model allows us to extend
previous MHD studies for coronal winds from young stars to spine
jets from Active Galactic Nuclei surrounded by disk-driven
outflows. The outflows are thermally driven and magnetically or
thermally collimated. The collimation depends critically on an
energetic integral measuring the efficiency of the magnetic
rotator, similarly to the non relativistic case. It is also shown
that relativistic effects affect quantitatively the depth of the
gravitational well and the coronal temperature distribution in the
launching region of the outflow. Similarly to previous analytical
and numerical studies, relativistic effects tend to increase the
efficiency of the thermal driving but reduce the effect of
magnetic self-collimation.}

\maketitle


\section{Introduction}

The formation of relativistic jets around compact objects and
Active Galactic Nuclei (AGNs) is one of the most intriguing and
yet not fully understood astrophysical phenomena
(\cite{Ferrari98}, \cite{Mirabel&Rodriguez98}, \cite{Mirabel03}).
In those jets, velocities reach a fraction of the speed of light
with the corresponding Lorentz factor ranging from values $\gamma
\sim 2 - 10$ in Seyfert Galaxies and radio loud AGNs (Piner et
al. 2003, \cite{UrryPadovani95}) up to the inferred values
$\gamma=10^3$ in GRBs; AGN jets are also characterized by the
rather narrow opening angles of a few degrees
(Biretta et al. 2002, Tsinganos \& Bogovalov 2005).

MHD models for coronal or disk-jets rely on the basic idea that
the gravitational energy of the central object is transferred to
the accreting plasma which via a collimation mechanism then
produces the jet. This energy released by accretion increases with
the mass of the central object, a fact which may explain the wide
variety of the powerful jets observed. Several analytical and
numerical efforts have been invested to investigate the mechanisms
of jet acceleration and collimation. The formation of collimated
jets seems to be closely related to the presence of large scale
magnetic fields (e.g., \cite{Gabuzda03}) and the existence of a
gaseous disk and/or a hot corona around the central object
(\cite{KoniglPudritz00}, \cite{Livio02}).

{ Regarding the {\it energy source} of jets, it is usually
assumed that at their base they are powered either by a spinning
black hole (Blandford \& Znajek 1977, \cite{Reesetal82},
\cite{Begelmanetal84}), or, by  the surrounding accretion disk 
(Miller \& Stone 2000).  
{}Furthermore, they are plausibly Poynting flux dominated (\cite{Sikora05}) 
with their central spine hydrodynamically dominated (\cite{Melianietal04}) 
and their plasma composed by protons-electrons or by electron-positron pairs. }
 
Regarding the {\it collimation} of the outflow, this is likely to
be due mainly to the hoop stress resulting from the toroidal
magnetic field generated by the rotation of the source (Bogovalov
1995). Recent VLBI observations suggest that the direction of the
magnetic field vectors is transverse to the jet axis. This is the
case in BL Lac objects, e.g., in Mrk 501  (\cite{Gabuzda03}), or,
in quasars where the central faster part of the jet is
characterized by toroidal magnetic fields (\cite{Asadaetal02}).
Magnetic self-collimation has been shown to be efficient in the
non relativistic context ({ {\cite{HeyvaertsNorman89}}, 2003},
\cite{Livio02}, \cite{HondaHonda02}, \cite{TsinganosBogovalov02}).
In the relativistic limit however it is
slower due to the decollimating effect of the electric force and
the higher inertia of the flow, but still possible (Vlahakis \&
K\"onigl 2003, 2004). Alternatively, collimation in relativistic
jets may be due to the external pressure of a surrounding slower
and easily collimated disk wind, \cite{BogovalovTsinganos05}.

Magnetized non relativistic jets from extended accretion disks
were first modelled analytically by  Blandford \& Payne (1982),
wherein the plasma acceleration relies on the magnetic extraction
of angular momentum and rotational energy from the underlying cold
Keplerian disk. This energy is channeled along the large-scale
open magnetic fieldlines anchored in the corona or the rotating
disk. The ionized fluid is forced to follow the fieldlines and to
rotate with them while it is  magnetocentrifugally accelerated if
the angle between the poloidal magnetic field and the disk is less
than $60^{\circ}$. \cite{CaoSpruit94} showed that in the relativistic
case this condition is less severe and that close to the black
hole the magnetocentrifugal acceleration may be efficient even at
higher angles. Analytical, radially self-similar disk-wind
solutions were extended to special relativistic cold winds in Li
et al. (1992), and Contopoulos (1994), by neglecting gravity to
allow the separation of the variables. Thermal effects were
introduced into these relativistic models by Vlahakis \& K\"onigl
(2003) to analyze the formation of a relativistic flow from hot
magnetized plasmas, showing that such solutions could be applied
to Gamma Ray Bursts wherein the flow is thermally driven at the
base. However, most of the acceleration is of magnetic origin and
there is an efficient conversion of Poynting to kinetic flux of
the order of 50\%. They also applied this disk wind solution to
AGN jets (\cite{VlahakisKonigl04}) showing that they could trace
the observed parsec scale expansion of the wind. Another approach
to solve the relativistic MHD equations for outflows around black
holes is to solve numerically the transfield equation in the
force-free limit (\cite{Camenzind86a}),
a study further developed in GRMHD by using first a Schwarzschild
metric and then extending it to a Kerr metric (\cite{Fendt97}).

Radial self-similarity is usually used in disk-wind models due to
the complexity of the non linear system of MHD equations. However,
such solutions cannot describe the flow close to the rotational
axis where they become singular. On the other hand, meridional
self-similarity provides a better alternative to study the outflow
close to the symmetry and rotation axis of the central corona. In
the central part where the thermal energy is rather high, the wind
may be thermally driven. Spherically symmetric relativistic
hydrodynamical models have been proposed to study the formation of
such outflows (\cite{Michel72, Das99, Melianietal04}). Those
models are restricted to the case where the magnetic effects in
the acceleration are negligible. In these models, a wind forms in
the hot corona because of the internal shock maintained by the
centrifugal barrier (\cite{Chakrabarti89, Das01}), or by the
pressure induced via a first order  Fermi mechanism
(\cite{Das99}). 

An important alternative to analytical models are numerical
simulations. In the special relativistic domain, simulations have been
presented for coronal winds (Bogovalov \& Tsinganos 1999,
Tsinganos \& Bogovalov 2002), and in the general relativistic
domain for disk winds (e.g.  {\cite{Koideetal99}}, 2001) to
model the formation and collimation of relativistic outflows in
the vicinity of black holes. The difficulty for relativistic
outflows in a single-component model to be collimated led
\cite{BogovalovTsinganos05} to propose a two-component model
wherein a relativistic central wind is collimated by a surrounding
non relativistic disk-wind. Shocks may also develop as the
disk-wind collides and collimates the inner relativistic wind.
Note that all such simulations are performed by using
time-dependent codes. Analytical models conversely have presented
more sophisticated steady solutions of outflows to be used as
initial conditions in more complex simulations, albeit sacrificing
freedom on the chosen boundary conditions.

In this article we present an extension of the non relativistic
meridionally self-similar solutions (\cite{SautyTsinganos94},
hereafter ST94) to
the case of relativistic jets emerging from a spherical corona
surrounding the central part of a Schwarzschild black hole and its
inner accretion disk. 

 We will not discuss here the origin of the plasma, assuming that 
it can come from, e.g., the accretion disk or pair creation.
{}Furthermore, only the outflow process is considered, with 
the base of the corona placed at a few Schwarzschild radii, just
above the so called separating surface (\cite{Takahashietal90}).

Attention is also given to the contribution of the different
mechanisms, hydrodynamic and magnetic, to the acceleration and
collimation of the outflow, as in previous papers of this series
(\cite{STT99}, 2002, 2004, hereafter STT99, STT02 and STT04).
This is also a way to extend to 2D outflows, thermally driven,
spherically symmetric (1D) wind models (\cite{Priceetal03},
\cite{Melianietal04}).

In the following two sections, the basic steady axisymmetric GRMHD
equations are presented using a 3+1 formalism  (Sect. 2), together
with their integrals.  The assumptions leading to the self-similar
model are presented in Sect. 3. In Sect. 4,
the analytical expressions of the model together with the derivation
of an extra free integral controlling the efficiency of the
magnetic rotator, as in STT99, are given. In Sect. 5, an asymptotic
analysis of the solutions is performed as well as the link to the
boundary conditions in the source. Sect. 6 is devoted to a
parametric study of various solutions to emphasize the main
difference obtained with relativistic flows. We discuss the
acceleration and collimation of these new solutions (Sect. 7) and
compare them with the non relativistic model in Sect. 8.  In the
last Sect. 9 we summarize our results and shortly outline their main
astrophysical implications. The
confrontation of the present model with observed jets from radio
loud extragalactic jets, such as those associated with FRI and FRII 
sources, is postponed to a following paper, as it involves
special techniques for constraining the parameters by using
observational data and a specific iterative scheme to use the
model.

\section{Basic equations}
In this section we briefly present, in order to establish
notation, the governing equations for magnetized fluids in the
background spacetime of a Schwarzschild black hole and also the
corresponding MHD integrals.
\subsection{The 3+1 formalism for steady flows}
\subsubsection{Schwarzschild metric}

The gravitational potential due to the matter outside the black
hole is assumed to be negligible. In Schwarzschild coordinates
($ct$, $r$, $\theta$, $\varphi$) the background metric is written
as,
\begin{equation}
{\rm d}s^2 = -{h_{}}^2 c^2 {\rm d}t^2 + \frac{1}{h_{}^2}
{\rm d}r^2+r^2 {\rm d}^2\theta+r^2 \sin^2{\theta} {\rm d}\varphi^2
\,,
\end{equation}
where
\begin{equation}
h_{} = \sqrt{1 - \frac{2 {\cal G}{\cal M}_{\bullet}}{c^2 r}}=
\sqrt{1-\frac{r_{\rm G}}{r}}\,,
\end{equation}
is the redshift factor induced by gravity at a distance $r$ from
the central black hole of mass ${\cal M}_{\bullet}$, expressed in
terms of the Schwarzschild  radius $r_{\rm G}=2{\cal G}{\cal
M}_{\bullet}/c^2$. Note that the time line element or the lapse
function is usually denoted by $h_{0}$ or $\alpha$. In this paper,
for further convenience and simplification we have used the symbol
$h$.

In the following we find convenient to use a 3+1 split of space-time,
following the usual
approach of MHD flow treatment in general relativity
(\cite{ThorneMcDonald82, Thorneetal86, MobarryLovelace86, Camenzind86a}).
The 3+1 approach allows to obtain equations similar to
the familiar classical equations. We write all quantities in 
the FIDucial Observer frame of reference, known as FIDO,  which corresponds
to observers in free fall around the Schwarzschild black hole. For 
the FIDO, space time is locally flat.

\subsubsection{Particle conservation}

The equation of conservation of particles
$(n u^a)_{;a} = 0$  in the 3+1 formalism is
\begin{equation}
\nabla\cdot (h_{}\gamma n\vec{V})= 0 \label{mass}
\,.\end{equation} Here $n$ is the proper number density of
particles, $\vec{V}$ is the fluid three-velocity as measured by FIDOs, 
$u^a = (\gamma c,\gamma \vec{V})$ and
\begin{equation}
\gamma=\left(1-\vec{V}^2/c^2\right)^{-1/2}
\,,\end{equation}
is the Lorentz factor.

\subsubsection{Maxwell's equations and Ohm's law}

Maxwell's equations written in the 3+1 formulation (Thorne \&
MacDonald 1982, Breitmoser \& Camenzind 2000) are\footnote{
Apparently, Eqs. (\ref{champsE}-\ref{fara}) have some differences
with their equivalent forms in a globally flat spacetime. The
$h_{}$ appearing in Eqs. (\ref{rotE}) and  (\ref{fara}) results
from formulating the laws of differentiation on a
``per-unit-t-base'' while the vectors $\vec{E}$ and $\vec{B}$ are
defined on a ``per-unit-FIDO-time'' (\cite{Thorneetal86}).}

\begin{eqnarray}
\nabla\cdot\vec{E}=4\pi\rho_{e}\,,\label{champsE}\\
\nabla\cdot\vec{B}=0\,,\label{dipoleM}\\
\nabla\times(h_{}\vec{E})=0\,,\label{rotE}\\
\nabla\times(h_{}\vec{B})=\frac{4\pi h_{}}{c}\vec{J}_{e}\label{fara}
\,,\end{eqnarray}
where ($\vec{E}, \vec{B}$) is the electromagnetic field and
($\rho_{e}, \vec{J}_{e}$) the associated charge and current densities.
Ohm's law for a plasma of high conductivity is
\begin{equation}
\vec{E} = - \frac{\vec{V}\times\vec{B} }{c} \label{conductivite 2}
\,.\end{equation}

\subsubsection{Euler Equation}

Euler's equation is obtained by projecting the conservation of the
energy-momentum tensor, $T^{a b}_{;b} =0$, onto the spatial
coordinates ($a=1,2,3$) and combined with Maxwell's equations
(see  \cite{BreitmoserCamenzind00}, or, for an expression closer
to ours, albeit restricted to special relativity Goldreich \&
Julian 1970, Appl \& Camenzind 1993, Heyvaerts \& Norman 2003),
\begin{eqnarray}
\gamma n (\vec{V}\cdot\nabla)\left(\frac{\gamma w\vec{V}}{c^2}\right)
&=&
-\gamma^2 n w \nabla\ln h_{}
- \nabla P
\nonumber\\
&&+\rho_{e}\vec{E}
+\frac{\vec{J}_{e} \times\vec{B}}{c}
 \label{Euler}
\,,\end{eqnarray} where $P$ and $w$ are the pressure and enthalpy
per particle, respectively. { The form of Eq. (\ref{Euler}) is equivalent
to that in Mobarry \& Lovelace (1986).}

\subsubsection{Thermodynamics}

The first law of thermodynamics is obtained by projecting the
conservation of the energy-momentum tensor  along the fluid
four-velocity, $ u_a T_{;b}^{ab}=0$. In fact, for ideal MHD fluid
the corresponding contribution of the electromagnetic field is
null due to the assumed infinite conductivity. Thus, only the
thermal energy affects the variation of the proper enthalpy of the
fluid,
\begin{equation}
\label{first_principe}
n\vec{V}\cdot\nabla w=\vec{V}\cdot\nabla P
\,.
\end{equation}

\subsection{Integrals of axisymmetric MHD outflows}

Assuming axisymmetry of the plasma flow allows us to reduce the
number of differential equations by integrating some of them and
thus obtaining conserved quantities along the streamlines. We
follow the notations of Tsinganos (1982).

{}From Eqs. (\ref{dipoleM}) and (\ref{mass}) we can introduce a magnetic
flux function $A$,
\begin{equation}
\vec{B}_{\rm p}=\nabla\times\left(\frac{A}{r\sin\theta}
\vec{e}_{\varphi}\right)=\frac{\nabla A}{\varpi}\times\vec{e}_{\varphi},
\,,
\label{Adef}
\end{equation}
and a stream function $\Psi$ which gives the particle flux,
\begin{equation}
4\pi h_{}\gamma n\vec{V_{\rm p}}=\nabla\times
\left(\frac{\Psi}{r\sin\theta}\vec{e}_{\varphi}\right)
=\frac{\nabla\Psi}{\varpi}\times\vec{e}_{\varphi}, \label{Psidef}
\,,\end{equation} where the subscript $\rm p$ denotes the poloidal
components and $\varpi=r\sin\theta$.

{}From Eq. (\ref{rotE}), we can define an electric potential associated to the
electric field, $\vec{E} = (\nabla\Phi) / h_{}$. Thus, the previous
equation and axisymmetry imply $E_{\varphi} = 0$.
In addition from the flux freezing condition
(Eq. \ref{conductivite 2}) and Eqs. (\ref{Adef}) and (\ref{Psidef})
we get that  $\Psi$ is constant on surfaces of
constant $A$ on which the corresponding streamlines and fieldlines are
roped, $\vec{V}_{\rm p} \parallel \vec{B}_{\rm p}$. It follows that
${\rm d}\Psi/{\rm d}A= \Psi_A$ is a function of $A$ and we can write
\begin{equation}
\vec{V_{\rm p}}=\frac{\Psi_A}{4\pi h_{}\gamma n}\vec{B_{\rm p}}
\,.\end{equation}

Since $\vec{B}_{\rm p}\cdot \nabla\Phi/h_{}=\vec{B}_{\rm p}\cdot
\vec{E} =\vec{B} \cdot \vec{E}=0$, the surfaces of constant
electric potential are also surfaces of constant magnetic flux, so
$\Phi =\Phi(A)$. Thus $\Omega = -c \, {\rm d}\Phi/{\rm d}A$ is
also a function of $A$. {}From Eq. (\ref{conductivite 2}) the
toroidal components $V_{\varphi}$ and $B_{\varphi}$ are related
\begin{equation}
V_{\varphi} = \frac{\Psi_A}{4\pi h_{}\gamma n}
B_{\varphi}+\frac{\varpi\Omega}{h_{}} \,.\end{equation} This is
called the isorotation law because each stream/fieldline rotates
rigidly with an angular speed $\Omega$ corresponding to the
angular speed $\Omega$ of the footpoints of this poloidal
stream/fieldline.

The azimuthal component of the momentum equation yields the
conservation of the total specific angular momentum,
\begin{equation}
L = \gamma\frac{w}{c^2}\varpi V_{\varphi}-
h_{} \frac{\varpi B_{\varphi}}{\Psi_A} \label{L}
\,.
\end{equation}

The generalized Bernoulli integral (including rest mass)
\begin{equation}
{\cal E} =h_{} \gamma w
-h_{}\frac{\varpi \Omega}{\Psi_A}B_{\varphi}  \label{E 2}
\,,\end{equation}
may be obtained by integrating the equation of motion along each streamline.
The first part of the r.h.s.  represents the hydrodynamical
energy flux transported by the fluid while the second part corresponds to
the Poynting flux.

We have obtained the usual four integrals of motion $\Psi_A$,
$\Omega$, ${\cal E}$, $L$ (\cite{HeyvaertsNorman03}) that are
constant along a fieldline for a stationary and axisymmetric
plasma. They can be used to find algebraic relations between  the
Lorentz factor, the toroidal velocity and the toroidal magnetic
field. Defining the poloidal ``Alfvenic'' number $M$
(\cite{Michel69}, \cite{Camenzind86b},
\cite{BreitmoserCamenzind00})
\begin{equation}
M_{}^2
=\frac{4\pi h_{}^2 n w \gamma^2 V_{\rm p}^2} {B_{\rm p}^2 c^2}
=\frac{w\Psi_A^2}{4\pi n c^2} \label{Mach number 1}
\,,\end{equation}
and the cylindrical distance in units of the light cylinder distance
(although in our case it is not a cylinder),
\begin{equation}
x=\frac{\varpi \Omega}{c h_{}}
\,,\label{xdef}
\end{equation}
we find,
\begin{eqnarray}
V_{\varphi}&=&\frac{c}{x}\left[\frac{M_{}^2 x_{\rm A}^2-x^2 h_{}^2
\left(1-x_{\rm A}^2\right)}{M_{}^2-h_{}^2
\left(1-x_{\rm A}^2\right)}\right],\label{vphi}\\
B_{\varphi}&=&-\frac{{\cal E}\Psi_A}{c x}
\left[\frac{x^2- x_{\rm A}^2}{M_{}^2-h_{}^2(1-x^2)}\right],\label{bphi}\\
h_{} \gamma w&=&{\cal E}\left[\frac{M_{}^2-h_{}^2 (1-x_{\rm
A}^2)}{M_{}^2-h_{}^2(1-x^2)}\right] \label{ag} \,.\end{eqnarray}
The quantity $x_{\rm A}^2$,
\begin{equation}\label{Lambda}
x_{\rm A}^2=\frac{\Omega L}{{\cal E}}
\end{equation}
is a measure for the amount of energy carried by the
electromagnetic field. It is a measure of the energy flux of the
magnetic rotator in units of the total energy flux
(\cite{BreitmoserCamenzind00}).

The MHD equations possess a well known singularity at the Alfv\'en
surface  where the denominator of Eqs. (\ref{vphi}), (\ref{bphi})
and (\ref{ag}) vanishes. Then, the numerators should vanish
simultaneously to ensure a regular behavior, implying at the
Alfv\'en point,
\begin{eqnarray}
\left. x^2\right|_{\mbox{Alfv\'en}}
=\left(\frac{\varpi_{\rm A}\Omega}{h_{\star}c}\right)^2=x_{\rm A}^2
\,,
\label{Lambdastar}\\
\left. M_{}^2\right|_{\mbox{Alfv\'en}}=
h_{\star}^2\left(1-x_{\rm A}^2 \right)
=M_{\rm A}^2
\label{Mstar}
\,.\end{eqnarray}
Using Eq. (\ref{Mach number 1}) we find
\begin{equation}
V^2_\star = \frac{B_\star^2 c^2}{4\pi\gamma_{\star}^2 n_{\star}w_{\star}}
\,,\end{equation}
where the subscript $\star$ denotes quantities evaluated at the Alfv\'en
point along the polar axis.  Note that the position of the
Alfv\'en  surface is shifted with respect to the classical case because
 of the lapse function  $h_{} $ and the existence of the light cylinder $x=1$.

In the Newtonian limit, Eqs. (\ref{Lambda}) and (\ref{Lambdastar}) give
$L=m\varpi_{\rm A}^2\Omega$ where $m=w/c^2$ is the
particle mass.

\section{Construction of the meridionally self-similar model}

Our goal in this section is to find semi-analytical solutions of the
$r-$ and $\theta-$ components of the Euler Eq. (\ref{Euler}), by
means of  separating the variables $r$ and $\theta$.
In order to facilitate the analysis it is convenient to use dimensionless
quantities normalizing at the Alfv\'en radius along the polar axis.
 Using the notations introduced in ST94 we define a dimensionless 
radius $R$ and   magnetic flux function $\alpha$,
\begin{equation}
R=\frac{r}{r_\star}\,, \qquad
A= \frac{r_\star^2 B_\star}{2}\alpha
\,.
\end{equation}

We introduce two dimensionless parameters to describe the
gravitational potential. The {\it first} is  $\nu$ which represents
the escape speed at the Alfv\'en point along the polar axis
in units of $V_\star$,
\begin{equation}
\nu=\frac{V_{{\rm esc,}\star}}{V_\star}
=\sqrt{\frac{2{\cal G}{\cal M}_\bullet}{r_{\star}V_{\star}^2}}
\label{Defnu}\,.\end{equation}

The {\it second} parameter\footnote{$\mu$
corresponds to the parameter $m$ used by Daigne \& Drenkhahn
(2002).} is the ratio of the Schwarzschild radius over the Alfv\'en
radius $r_\star$,
\begin{equation}
\mu=\frac{r_{\rm G}}{r_\star} = \frac{V_{{\rm esc,}\star}^2}{c^2}
\label{Defmu}\,.
\end{equation}
which is also the escape speed in units of the speed of light.

Combining Eqs. (\ref{Defnu}) and (\ref{Defmu}) we get a condition that
restrict the parametric space to
\begin{equation}
\frac{\mu}{\nu^2}=\frac{V_{\star}^2}{c^2}<1
\,.\end{equation}

\subsection{Separation of the variables}

\subsubsection{Magnetic flux and ``Alfv\'enic'' number}

As $\alpha=0$ on the rotational axis and we are interested on the
central component of the jet, we assume that the cross section area of a
given magnetic flux tube can be expanded to first order in $\alpha$,
\begin{equation}
S(R,\alpha)=\pi\varpi^2=\pi r_\star^2 G^2(R) \alpha
\,.
\end{equation}
Normalizing at the Alfv\'en surface, we choose $G(R=1)=1$ such that
$G$ is the cylindrical radius in units of the Alfv\'enic cylindrical
radius. Thus $G(R)={\varpi}/{\varpi_{\rm A}}$ with
$\varpi_{\rm A}=r_\star \sqrt{ \alpha}$.

This is equivalent to assume, as in ST94, that the magnetic flux
function $\alpha(R,\theta)$ has a dipolar latitudinal dependence,
\begin{equation}
\alpha=\frac{R^2}{G^2(R)}\sin^2 \theta
\,.\end{equation}

We also introduce the expansion factor $F$
\begin{equation}
{}F=\left. \frac{\partial \ln{\alpha}}{\partial \ln{R}}\right|_\theta
=2-\frac{{\rm d}\ln{G^2}}{{\rm d}\ln{R}}
\label{F-G}
\,.\end{equation}

In addition, we assume that the surfaces of constant
``Alfv\'enic'' number are spheres, such that,
\begin{equation}
M_{}^2(R,\alpha)=M^2(R)
\,, \label{Mach number 2}\end{equation}
with
$M^2(R=1) = h_{\star}^2$.

\subsubsection{Pressure}

The pressure dependence is obtained by making a first order expansion in $\alpha$
\begin{equation}
P=P_0 + \frac{1}{2}{\gamma_{\star}}^2n_{\star}\frac{w_\star}{c^2}
V_ {\star}^2\Pi(R)(1+\kappa\alpha) \label{pressure}
\,,\end{equation} with $P_0$, $\kappa$ constants and $\Pi (R)$ a
dimensionless function.

\subsubsection{Free integrals}

Combining Eqs. (\ref{Mach number 1}) and (\ref{Mach number 2})  we deduce that
\begin{equation}
\label{eq-n}
\frac{4\pi n c^2 M^2}{w}=\Psi_{A}^2
\,.\end{equation}
Expanding the r.h.s. to first order in $\alpha$, we find
\begin{equation}
\label{psi_A}
 \Psi_A^2=4\pi
c^2\frac{n_{\star}h_{\star}^2}{w_{\star}}(1+\delta\alpha)
\,,\end{equation} where $\delta$ is a free parameter describing
the deviations from spherical symmetry of the ratio number
density/enthalpy and not of the density itself  as in ST94.

Similarly we expand  $L\Psi_A$
\begin{eqnarray}
L\Psi_A=h_{\star}\lambda B_{\star}r_{\star} \alpha \label{LPSI}
\,,
\end{eqnarray}
where $\lambda$ is a constant measuring rotation.

{}Finally  we choose for $\Omega$ a form similar to the
one in ST94
\begin{eqnarray}
\Omega=\lambda h_{\star}
\frac{V_{\star}/r_{\star}}{\sqrt{1+\delta\alpha}}
\label{Omega}
\,.
\end{eqnarray}

{}From Eq. (\ref{xdef}) we can now express $x^2$ in terms of
$\alpha$ and $G(R)$,
\begin{equation}
x^2 =\lambda^2 \frac{V_\star^2}{c^2} \frac{h_{\star}^2}{h_{}^2} G^2
\frac{\alpha}{ 1 + \delta\alpha}
=\frac{\mu\lambda^2}{\nu^2} \frac{h_{\star}^2}{h_{}^2} G^2
\frac{\alpha}{ 1 + \delta\alpha}
\,.\label{xa_model}\end{equation}
Similarly, from Eq. (\ref{Lambdastar}) we find
\begin{equation}
x_{\rm A}^2 =\lambda^2 \frac{V_\star^2}{c^2}\frac{\alpha}{ 1 + \delta\alpha}
=\frac{\mu\lambda^2}{\nu^2}\frac{\alpha}{ 1 + \delta\alpha}
\,,\label{xa_model}\end{equation}
which gives from Eq. (\ref{Lambda}) the form of the Bernoulli integral,
\begin{equation}
{\cal E}=h_{\star}\gamma_{\star}w_{\star}
\label{Ecritique}
\,.\end{equation}
Note that the parameter $\lambda$ in Eq. (\ref{Omega})
is the same constant as in Eq. (\ref{LPSI}) because the energy ${\cal E}$
must be equal to its hydrodynamic part $h_{\star}\gamma_{\star}w_{\star}$
on the rotational axis where the Poynting flux vanishes.

\subsection{Electric force}

Conversely to the non relativistic limit, we cannot neglect the
charge separation and the presence of the electric field. From the
previous assumptions, we can calculate the electric force
$\rho_{e}\vec{E}$ which has the  following two components
\begin{eqnarray}\label{Jet_1_ElectricF}
\rho_e  E_r & = &\frac{B_\star^2}{4 \pi r_{\star} G^4 }
\left\{\frac{h_{\star}^2}{h_{}}
\frac{F}{2} x_{\rm A}^2 G^2 \sin^2\theta \right.\nonumber\\
&&\left.\left[ \frac{h_{}^2}{4}
\left( 2 \frac{{\rm d} F}{{\rm d} R} + \frac{F^{2}}{R}
+ 2 F
-\frac{8}{h_{}^2 R} \right) \right.\right.\nonumber\\
&&\left.+ \frac{1}{R \left(1 + \delta \alpha\right)} \left(h_{}^2\frac{F^2}{4}
- 1\right)\right]\nonumber\\
&&+\left.\frac{h^2_{0\star}}{h_{}} \frac{F}{2}
\lambda^2 \frac{V^2_{\star}}{c^2} R \sin^2\theta
\frac{2 + \delta \alpha}{(1 + \delta \alpha)^{2}}\right\}\,,
\\
\rho_e  E_\theta & = &
\frac{B_{\star}^2}{4 \pi r_{\star} G^4} \left\{ \left(\frac{
h_{\star}}{h_{}}\right)^2 x_{\rm A}^2 G^2 \sin\theta \cos\theta
\right.\nonumber\\
&&\left.\left[ \frac{h_{}^2}{4}
\left( 2 \frac{{\rm d} F}{{\rm d} R} + \frac{F^{2}}{R}
+ 2 F
-\frac{8}{h_{}^2 R} \right) \right.\right.\nonumber\\
&&\left.+ \frac{1}{R \left(1 + \delta \alpha\right)} \left(h_{}^2
\frac{F^2}{4} - 1\right)\right]\nonumber\\
&&+ \left. \left(\frac{h_{\star}}{h_{}}\right)^2 \lambda^2
\frac{V^2_{\star}}{c^2} R\sin\theta \cos\theta
\frac{2 + \delta \alpha}{(1 + \delta \alpha)^{2}}\right\}
\,.
\end{eqnarray}
Note that in this form of the expressions of the forces the
variables are not separable.

\subsection{Non relativistic rotation}

Contrary to the non relativistic case and in order to separate the
variables $(R, \theta)$ in the $r-$ and $\theta-$ components of
the Euler Eq. (\ref{Euler}) we need some further assumptions.
Basically we expand these equations with respect to $\theta$.

We suppose that the rotational speed of the fluid remains always
subrelativistic. In other words, we focus on streamlines
that never cross the
light cylinder such that the later does not affect the dynamics
($x\ll 1$, which implies $x_{\rm A}\ll 1$). Of course this reduces
the domain of validity of the solutions to the vicinity of the
rotational axis because
$x_{\rm A}^2$ as it is given by Eq. (\ref{xa_model})
is sufficiently small only for relatively small $\alpha$. The
region of validity of our model depends on how small the parameter
$\lambda^2 V_\star^2 /c^2=\lambda^2 \mu / \nu^2$ is, though. The
requirement that the light cylinder lies further away from the
Alfv\'enic surface $(x_{\rm A}<1)$ constrains the parametric space
to $\delta>\lambda^2 \mu / \nu^2$. On the other hand, there is a
possibility that all streamlines never cross the light cylinder. This
happens when the equation
\begin{equation}\label{Jet_1_Surface_lumiere}
x=1 \Leftrightarrow
\frac{1}{\alpha}=\frac{\lambda^2\mu}{\nu^2}
\frac{G^2 h_{ \star}^2}{h_{}^2}-\delta
\end{equation}
cannot be satisfied for any $R$, or equivalently
$\delta > (\lambda^2\mu/ \nu^2 )
\left[G^2 {h_{ \star}^2}/{h_{}^2} \right]_{\max}$.
Weak rotation (small $\lambda$) or significant deviation of the particle flux
from spherical symmetry (high $\delta$ which results in a fast decrease of
$\Omega$ as we move away from the rotation axis)
contribute to the validity of the above inequality.

A consequence of this is that the jet is thermally driven. Indeed the ratio
between the Poynting flux and the matter energy flux is,
\begin{equation}
\frac{ - h_{} \varpi \Omega B_\varphi /\Psi_A}{h_{} \gamma w}
=h_{}^2\frac{x^2-x_{\rm A}^2}{M_{}^2-h_{}^2(1-x_{\rm A}^2)}
\,.\end{equation} Thus, in the approximation $x\ll 1$ the
contribution of the Poynting flux is negligible in accelerating
the flow.

More generally, after expanding with respect to $x_{\rm A}^2$, we
neglect terms of the order $x_{\rm A}^2 \sin \theta $ or higher.
This is also consistent with our previous assumption of keeping
terms only up to $\alpha$ in the integrals. For example, the
expression of the azimuthal magnetic field, Eq. (\ref{bphi}),
becomes
\begin{eqnarray}\label{bphi1}
B_{\varphi}
&\approx & -\frac{\lambda B_\star}{G^2}
\frac{h_{\star}}{h_{}} \frac{G^2 h_{\star}^2-h_{}^2} {M^2-h_{}^2}R\sin\theta
\,.
\end{eqnarray}
Equivalently, after expanding all terms with respect to $\theta $
we neglect terms of the order of $\sin^3 \theta $ or higher.

As another example, the approximate form of the electric force is
\begin{eqnarray}
\label{rEr}
\rho_e  E_{r} \approx
\frac{B_\star^2}{4\pi r_{\star} G^4} \frac{h^2_{0\star}}{h_{}} F
\lambda^2 \frac{V^2_{\star}}{c^2} R \sin^2\theta
\,, \\
\label{rEQ}
\rho_e  E_{\theta} \approx
2\frac{B_{\star}^2}{4\pi r_{\star} G^4}
\left(\frac{h_{\star}}{h_{}}\right)^2\lambda^2
\frac{V^2_{\star}}{c^2}R\sin\theta\cos\theta
\,,
\end{eqnarray}
where we have further approximated $\frac{2 + \delta
\alpha}{\left(1 + \delta \alpha\right)^2} \approx 2$, since this
factor is multiplied with $x_{\rm A}^2 / \sin \theta \propto \sin
\theta $ and $\alpha \propto \sin^2 \theta$. Though this
approximation is consistent with the fact that we neglect any
terms of the order of $\sin^3 \theta $ or higher, it gives an
extra restriction.  Thus, the model applies only to the region
near the rotational axis where $\alpha \ll 1/ \delta$. We shall
calculate both inequalities ($x_{\rm A}^2 \ll 1$ and $\alpha \ll
1/ \delta$) {\sl a posteriori} in order to determine the regime
where the solution is valid.

  Note that in the very vicinity of the black hole $h \rightarrow 0$
and since the factor $h x$ does not vanish (see Eq. \ref{xdef}), 
$x$ becomes larger than unity,
as expected by the presence of the second light cylinder close to the horizon
(e.g., Takahashi et al. 1990).  
However, it is enough that the coronal base is at a distance of a few 
gravitational radii ($R \gapp 2 \mu$) to guarantee that $x\ll1$
at the base of the outflow (this condition is fulfilled in our
solutions). 
 
\section{Equations of the model}

\subsection{Expressions of the fields and the enthalpy}

The velocity and magnetic fields can now be written exclusively in terms of
unknown functions of $R$. For later convenience, as in ST94, we shall denote by
$N_B$, $N_V$ and $D$ the following quantities that appear in various 
components of the fields,
\begin{eqnarray}
N_B=\frac{h_{}^2 }{h_{\star}^2} - G^2\,,\\
N_V=\frac{M^2}{h_{\star}^2}-G^2\,,\\
D=\frac{h_{}^2 }{h_{\star}^2}-\frac{M^2}{h_{\star}^2}
\,.\end{eqnarray}

Thus,
\begin{eqnarray}
\label{Br}
B_r&=&\frac{B_\star}{G^2}\cos{\theta}\,,\\
\label{BQ}
B_{\theta}&=&- \frac{B_\star}{G^2}\frac{h_{}F}{2}\sin{\theta}\,,\\
\label{Bphi}
B_{\varphi}&=&-\frac{\lambda B_\star}{G^2}
\frac{h_{\star}}{h_{}}
\frac{N_B}{D} R\sin{\theta} \,,\\
\label{Vr}
V_r&=&\frac{V_\star M^2}{h_{\star}^2G^2}
\frac{1}{\sqrt{1+\delta\alpha}}
\left(\cos{\theta}
+\frac{\mu \lambda^2}{\nu^2}\frac{N_B}{D} \alpha
\right) \,,\\
\label{VQ}
V_{\theta}&=&-\frac{V_\star M^2}{h_{\star}^2G^2}
\frac{h_{}F}{2}\frac{1}{\sqrt{1+\delta\alpha}} \sin{\theta}
\,,\\
\label{Vphi}
V_{\varphi}&=&-\frac{h_{} }{{h_{\star}}}
\frac{\lambda V_\star}{G^2}
\frac{N_V}{D}
\frac{R\sin{\theta}}{\sqrt{1+\delta\alpha}}
\,.\end{eqnarray}
The electric field can be deduced from the flux freezing condition and the
above equations (see Eqs. \ref{rEr},  \ref{rEQ}).

Similarly, the enthalpy and the particle number density are given by
\begin{equation}
\label{hgw}
h_{}\gamma w=h_{\star} \gamma_\star w_\star
\left(1-\frac{\mu \lambda^2}{\nu^2}
\frac{N_B}{D} \alpha \right)
\,,\end{equation}
\begin{equation}
\label{hgn}
 h_{}\gamma n=h_{\star} \gamma_\star  n_\star
\frac{h_{\star}^2}{M^2}\left(1+\delta\alpha -\frac{\mu
\lambda^2}{\nu^2} \frac{N_B}{D} \alpha
\right) \,,\end{equation} where we used Eq. (\ref{psi_A}) to
deduce Eq. (\ref{hgn}), while the pressure is given by Eq.
(\ref{pressure}).

\subsection{Ordinary differential equations and numerical technique}

There are three equations given in Appendix \ref{appendixA}
that determine the  three unknown functions $\Pi(R)$, $F(R)$ and
$ M^2(R)$.  We recall that $G^2$ is related to $F$ through Eq. (\ref{F-G}).
Before discussing in detail the results  of the  parametric study
we outline the method for the numerical integration
of  Eq. (\ref{F-G}) and Eqs. (\ref{dPdR}) - (\ref{dFdR}).
We start integrating the equations from the Alfv\'en critical
surface. In order to calculate the toroidal components of the fields, i.e.
$N_B/D$ and $N_V/D=N_B/D-1 $,  we apply L'H\^opital's rule,
\begin{equation}
\left.\frac{N_B}{D}\right|_\star =\frac{h_{\star}^2(2- F_\star)
-\mu } {h_{\star}^2 p} \,, \quad
p=\left.\frac{1}{h_{\star}^2}\frac{{\rm d}  M^2}{{\rm d} R}
\right|_{\star}-\frac{\mu}{h_{\star}^2} \,.
\end{equation}

To avoid kinks in the fieldline shape, we need to satisfy
a regularity condition (\cite{HeyvaertsNorman89}). This means that Eq.
(\ref{dFdR}) should be regular at $R=1$. As in ST94 this extra
requirement is equivalent to ${\cal N}_F, D= 0$ which eventually
gives a third order polynomial equation for $p$
\begin{equation}
C_3 p^3+C_2 p^2+C_1 p+ C_0 = 0 \,,
\end{equation}
\begin{eqnarray}
C_0 &=& - \frac{\lambda^2}{2} \left( F_\star - 2 + \frac{\mu}{h^2_{0 \star}} \right)^2 \,,\\
C_1 &=& \lambda^2 \left( F_\star - 2 + \frac{\mu}{h^2_{0 \star}} \right) \,,\\
C_2 &=&\frac{1}{2}\lambda^2-\frac{h_{\star}^2}{8} F_{\star}^2
+\frac{1}{2}(1-\kappa \Pi_{\star}) + \lambda^2 \frac{\mu}{\nu^2}\,,\\
C_3 &=&h_{\star}^2\frac{F_{\star}}{4}\,.
\end{eqnarray}

Once we have determined the regularity conditions at the Alfv\'en point, we
integrate downwind and upwind and cross all the other existing
critical points as in the non relativistic case.

It is worth noticing that the shape of the streamlines $F_{\star}$  at $R=1$ is determined by
the regular crossing of the slow magnetosonic surface. We point out further
that, besides the free parameters listed at the beginning of the section,
solutions depend also on  $\Pi_{\star}$, i.e.  the pressure at the Alfv\`en
surface.  As in the classical case its value has been chosen such that the gas pressure is always
positive. More details on the numerical technique can be found in ST94 and STT02.

\subsection{The integral $\epsilon$}

As in ST94, it is possible to find a constant $\epsilon$ for all fieldlines.
This parameter $\epsilon$ has been used in ST94 and in the following papers
to classify the various solutions. We shall use a similar technique to
construct such a constant in the present model.

Eq. (\ref{first_principe}), after substituting $n$ from
Eq. (\ref{Mach number 1}) and using ${\vec{V}}
\cdot \nabla  \propto {\partial }/{\partial R}|_\alpha$
(derivative keeping $\alpha$ constant),
can be re-written as
\begin{eqnarray}
-8 \pi M^2 \left. \frac{\partial P}{\partial R} \right|_\alpha
&=&-\frac{\partial}{\partial R} \left.\left(
\frac{\Psi_A^2 w^2}{c^2}\right)\right|_\alpha
\nonumber\\
&=&\frac{\partial}{\partial R} \left.\left[
\frac{\Psi_A^2 ({\cal E}^2 - w^2)}{c^2}\right]
\right|_\alpha
\label{integral1}
\,,
\end{eqnarray}
where $\Psi_A^2 w^2$ is proportional to
the energy {\it per unit volume} of the fluid in the comoving frame, i.e.
reduced to the thermal
content. Thus $\Psi_A^2 ({\cal E}^2-w^2)$
in essence measures the variation between the total energy
and the thermal energy of the fluid.

By writing the term
\begin{eqnarray}
\frac{\Psi_A^2 w^2}{c^2}&=&\frac{\Psi_A^2 w^2 \gamma^2 (1-V_\varphi^2/c^2-V_p^2/c^2)}{c^2}
\nonumber \\
&=&\frac{\Psi_A^2}{c^2 h_{}^2} \left(h_{} \gamma w \right)^2
\left(1-\frac{V_\varphi^2}{c^2} \right) -\frac{M^4 B_p^2}{h_{}^2} \,,
\end{eqnarray}
and using Eqs.  (\ref{vphi}) and (\ref{ag}) we find
\begin{eqnarray}
\frac{\Psi_A^2 w^2}{c^2}=
\frac{\Psi_A^2 {\cal E}^2}{c^2 h_{}^2}
\left[ \frac{M^2-h_{}^2(1-x_{\rm A}^2)}{M^2-h_{}^2(1-x^2)} \right]^2
\nonumber \\
-\frac{\Psi_A^2 {\cal E}^2}{c^2 h_{}^2}
\left[ \frac{M^2 x_{\rm A}^2/x-xh_{}^2(1-x_{\rm A}^2)}{M^2-h_{}^2(1-x^2)} \right]^2
-\frac{M^4 B_p^2}{h_{}^2}
\,.
\end{eqnarray}
In the particular model that we examine, the form of the pressure
is $P=f_1(R) (1+\kappa \alpha)/8 \pi$. We also know the $\theta$
dependence in all quantities in the expression for $\Psi_A^2
w^2/c^2$, and after expanding with respect to $\sin^2\theta$ we
find $\Psi_A^2 ({\cal E}^2-w^2)/c^2=f_2(R) + f_3(R) \alpha$. Then
Eq. (\ref{integral1}) gives
\begin{eqnarray}
-M^2 \frac{{\rm d} f_1}{{\rm d}R}(1+\kappa \alpha)
= \frac{{\rm d} f_2}{{\rm d}R} + \frac{{\rm d} f_3}{{\rm d}R} \alpha\nonumber\\
\Leftrightarrow \left\{
\begin{array}{ll}
-M^2 {\rm d} f_1 = {\rm d} f_2 \\
-M^2 \kappa {\rm d}f_1 = {\rm d}f_3
\end{array}
\right.
\end{eqnarray}
Eliminating ${\rm d} f_1$ we get the integral $ f_3(R) - \kappa f_2(R)
= \epsilon =$const.
After substituting the expressions for $f_2(R)$ and $f_3(R)$, we arrive at
\begin{eqnarray}\label{varepsilon_full}
\epsilon =
\frac{M^{4}}{h_{\star}^4 R^2 G^2}
\left(\frac{F^2}{4} - \frac{1}{h_{}^2} - \kappa \frac{R^2}{h_{}^2G^2}\right)
-\frac{\left(\delta-\kappa\right)\nu^2}{h_{}^2 R}
\nonumber \\
+\frac{\lambda^2}{G^2 h_{\star}^2} \left(\frac{N_V}{D}\right)^2
+\frac{2 \lambda^2}{h_{}^2}\frac{N_B}{D}
\,,
\end{eqnarray}
where $\epsilon$ is a constant, the same for all fieldlines.

Similarly to what was done in ST99, we can calculate this constant at
the base of the flow $R_o$
assuming the poloidal velocity is negligible there [$M(R_o)\approx
0$]. Let's express $\epsilon / 2\lambda^2$  in terms of
the conditions at the source boundary,
\begin{equation}
{\epsilon  \over 2\lambda^2} =\frac{{\cal E}_{{\rm R},o}
+ {\cal E}_{{\rm Poynt.},o}+ \Delta {\cal E}_{\rm G}^* }
{ {\cal E}_{\rm MR}}
\,,
\end{equation}
where ${\cal E}_{\rm MR}=h_{}^2 L\Omega $ is the energy of the magnetic rotator,
${\cal E}_{\rm Poynt.}=- h_{} \varpi \Omega B_\varphi /\Psi_A$ is the Poynting flux
and
\begin{equation}
{\cal E}_{{\rm R},o}=\frac{\cal E}{c^2}\frac{V_{\varphi, o}^2}{2}
\,,
\end{equation}
is the rotational energy per particle. It is proportional to the
specific rotational energy $V_{\varphi, o}^2/2$, with the factor
${\cal E}/{c^2}$ having the dimensions of a mass. Finally we have
\begin{equation}
\Delta {\cal E}_{\rm G}^*
= -\frac{\cal E}{c^2}\frac{\mu c^2}{2R_o}
\frac{\left(\delta-\kappa\right) \alpha}{1+\delta \alpha}
\,,
\end{equation}
a term similar to the nonrelativistic case where it measures the
excess or the deficit on a non polar streamline, compared to the
polar one, of the gravitational energy per unit mass which is not
compensated by the thermal driving (STT99). As in the classical
case, $\epsilon$ measures the efficiency of the magnetic rotator
to collimate the flow. Thus if $\epsilon>0$ we have an Efficient
Magnetic Rotator (EMR) where magnetic collimation may dominate,
while if $\epsilon<0$ we have an Inefficient Magnetic Rotator
(IMR) where collimation cannot be but of thermal origin.

\section{Asymptotic behaviour}

In the region far from the base where the jet attains its
asymptotic velocity, assuming it becomes cylindrical, the forces
in the radial direction become negligible, since the jet is no
longer accelerated. In the transverse direction, the following
four forces balance each other: the transverse pressure gradient,
$\vec{f}_{P}$, the total magnetic stress (magnetic pinching plus
magnetic pressure gradient) of the toroidal magnetic field
component, $\vec{f}_{B}$,  the centrifugal force, $\vec{f}_{C}$
and the charge separation electric force, $\vec{f}_{E}$,
\begin{equation}\label{Jet_1_force_Eq}
\vec{f}_{C} + \vec{f}_{B} + \vec{f}_{P} + \vec{f}_{E}= 0,
\end{equation}
The full expressions of these forces are given in Appendix
\ref{AppendixB}. In the asymptotic region, $\theta \sim 0$, they
can be written as follows in cylindrical coordinates,

\begin{eqnarray}\label{Jet_1_force_T}
f_C & = &\gamma^2 n \frac{w}{c^2} \frac{V_\varphi^2}{\varpi}
\nonumber \\& = &
 \frac{B_\star^2}{4 \pi G_\infty^4}
\frac{h_{ \star}^2\lambda^2}{M^2_\infty}{\varpi_\infty}
\left(\frac{{N_V}_{\infty}}{D_{\infty}}\right)^2
 \,,\\
f_{B} & = &-\frac{1}{4 \pi \varpi} \left( B_{\varphi}^2
+ \frac{1}{2}\frac{d B_{\varphi}^2}{d \varpi} \varpi \right)\nonumber \\& = &
- \frac{B_{\star}^2}{2 \pi G_{\infty}^4}
{h_{\star}^2} {\lambda^2} {{\varpi}_{\infty}}
\left( \frac{N_{B\infty}}{D_\infty}\right)^2 \,,\\
f_{P} & = &-\frac{{\rm d} P}{{\rm d} \varpi} =
-\frac{B_{\star}^2}{4 \pi  G_{\infty}^2}
\Pi_{\infty} \kappa {{\varpi}_{\infty}}  \,,
\label{Jet_1_force_T_P}\\
f_{E} &=&\rho_{e}E_\varpi=  \frac{B_{\star}^2}{2 \pi G_{\infty}^4}
{h_{\star}^2}\frac{\lambda^2\mu}{\nu^2}{\varpi_\infty}
\label{Jet_1_force_T_PP}
\,.
\end{eqnarray}

The centrifugal and electric forces have a decollimating effect on
the jet, while the pinching magnetic force collimates it because of our
choice on the current distribution. For asymptotically underpressured
jets where $\kappa>0$ and  $\Pi_{\infty}>0$ (or $\kappa<0$
and $\Pi_{\infty}<0$), the pressure increases away from
the polar axis which helps collimation. The opposite holds for
overpressured jets where $\kappa<0$ and $\Pi_{\infty}>0$ (or
$\kappa>0$ and $\Pi_{\infty}<0$).

Combining Eq. (\ref{Jet_1_force_Eq}) with Eqs. (\ref{Jet_1_force_T}) -
(\ref{Jet_1_force_T_PP}) we obtain,
\begin{eqnarray}
\label{Jet_1_Asy_Kappa_Conf}
\frac{\kappa}{2 \lambda^2} \Pi_{\infty}= \frac{h_{\star}^2}{G_\infty^2} \left[
\frac{1}{2 M_{\infty}^2} \left(\frac{{N_V}_{\infty}}
{D_{\infty}}\right)^2 + \frac{\mu}{\nu^2}
-\left(\frac{{N_B}_{\infty}}{D_{\infty}}\right)^2
\right]
\,.
\end{eqnarray}

The second equation controlling the asymptotic transverse force
balance is given by $\epsilon$, Eq. (\ref{varepsilon_full}).
{ In the asymptotic region, for a
cylindrically collimated jet $F_\infty \rightarrow 2$ and 
$\epsilon$ becomes},
\begin{equation}\label{Jet_1_Asy_Varepsilon_Conf}
\frac{\epsilon}{2 \lambda^2} = - \frac{\kappa}{2 \lambda^2}
 \frac{M^4_{\infty}}{h_{\star}^4 G^4_{\infty}}
+\frac{1}{2 h_{\star}^2 G^2_{\infty}} \left(\frac{{N_{V}}_{\infty}}
{D_{\infty}}\right)^2
+ \frac{{N_{B}}_{\infty}}{D_{\infty}}
\,.
\end{equation}
We notice that the asymptotic behaviour of the jet is described by the asymptotic pressure
$\Pi_{\infty}$ and the two parameters ${\epsilon}/{2\lambda^2}$ and ${\kappa}/{2\lambda^2}$.
These equations are similar to the classical model except for the decollimating
effect of the electric field and charge separation. Besides that note also that in $D$, $N_{V}$ and
$N_{B}$ the space curvature at the Alfv\'en critical surface also appears.

At the base of the wind the Alfv\'en number vanishes,
$M_{o}\rightarrow 0$, while the opening of the jet is weak,
$G^2_{o} \ll 1$. We can use this criterion to define the distance
$R_o$ where the outflow starts. Namely from the expression of
$\epsilon$, Eq. (\ref{varepsilon_full}), we get,
\begin{equation}\label{Jet_1_R_corona}
R_o = \frac{(\delta - \kappa)\nu^2 - \mu\epsilon} {2 \lambda^2 - \epsilon}\,.
\end{equation}
We see that in order to have acceleration ($R_o>\mu$) we must have
approximately $(\delta-\kappa )/2 > \mu \lambda^2 /\nu^2$ for $2
\lambda^2 - \epsilon>0$, extending the criterion found in the
classical regime ($\delta > \kappa$; the above expression reduces
to Eq. 14 of STT02 for $\mu \rightarrow 0$). In particular we note
that for $\epsilon >0$ or $\epsilon <0$ relativistic effects
enlarge or reduce the size of the sub-Alfv\'enic region,
respectively.

 Note that the definition of $R_o$ coincide with the so
called  separating  surface (see e.g. \cite{Takahashietal90}
for a cold plasma in a Kerr metric). Above
this  surface the plasma is 
outflowing. Below this surface other critical surfaces exist 
(e.g., \cite{BeskinKuznetsova00}) but remain
out of our consideration. If the plasma is created via pair 
production this may constrain the boundary conditions at the base of the 
flow. However we do not discuss the origin of the coronal plasma
in this paper.

\begin{figure*}
{\rotatebox{0}{\resizebox{8cm}{6cm}{\includegraphics{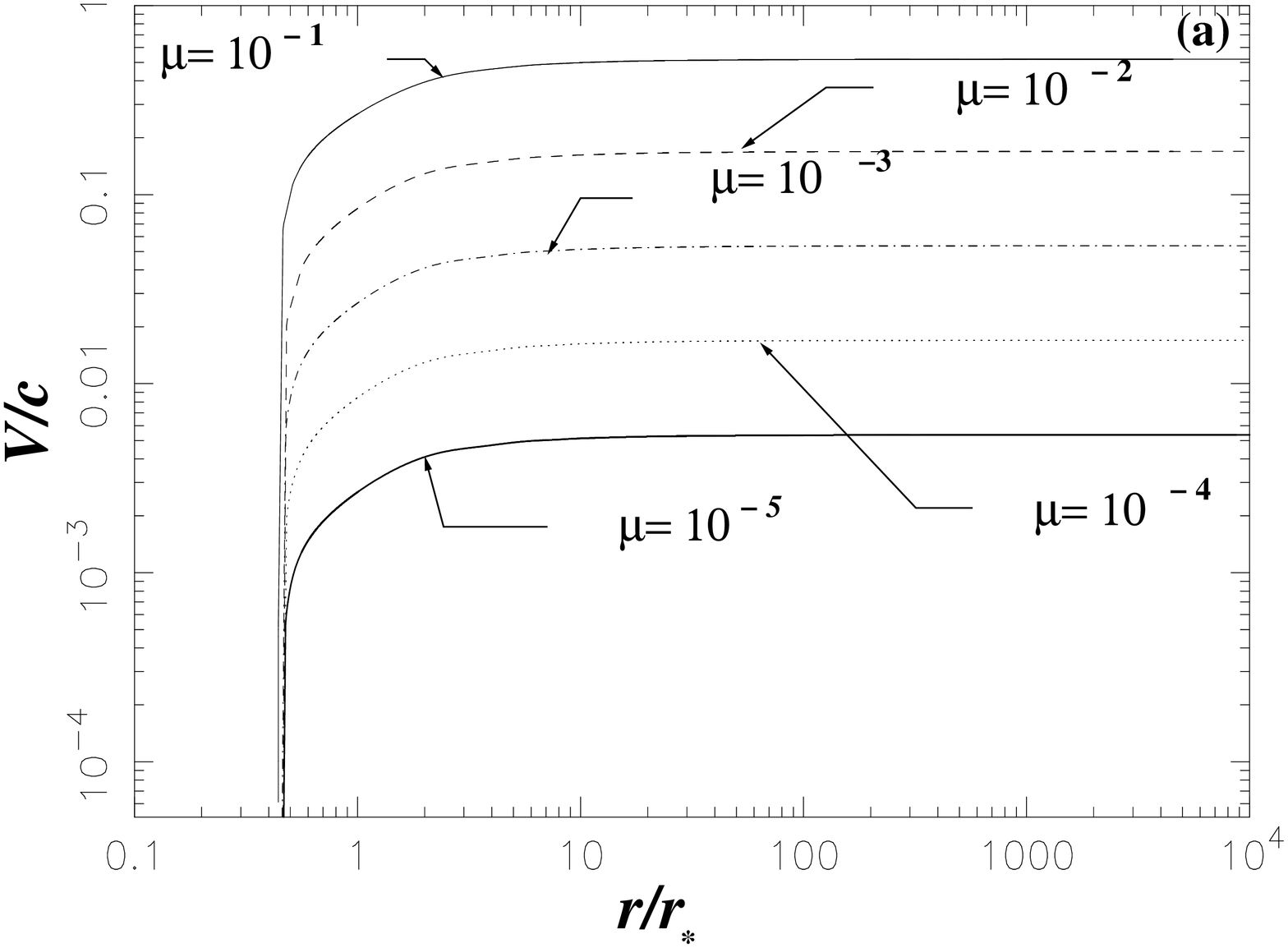}}}}
{\rotatebox{0}{\resizebox{8cm}{6cm}{\includegraphics{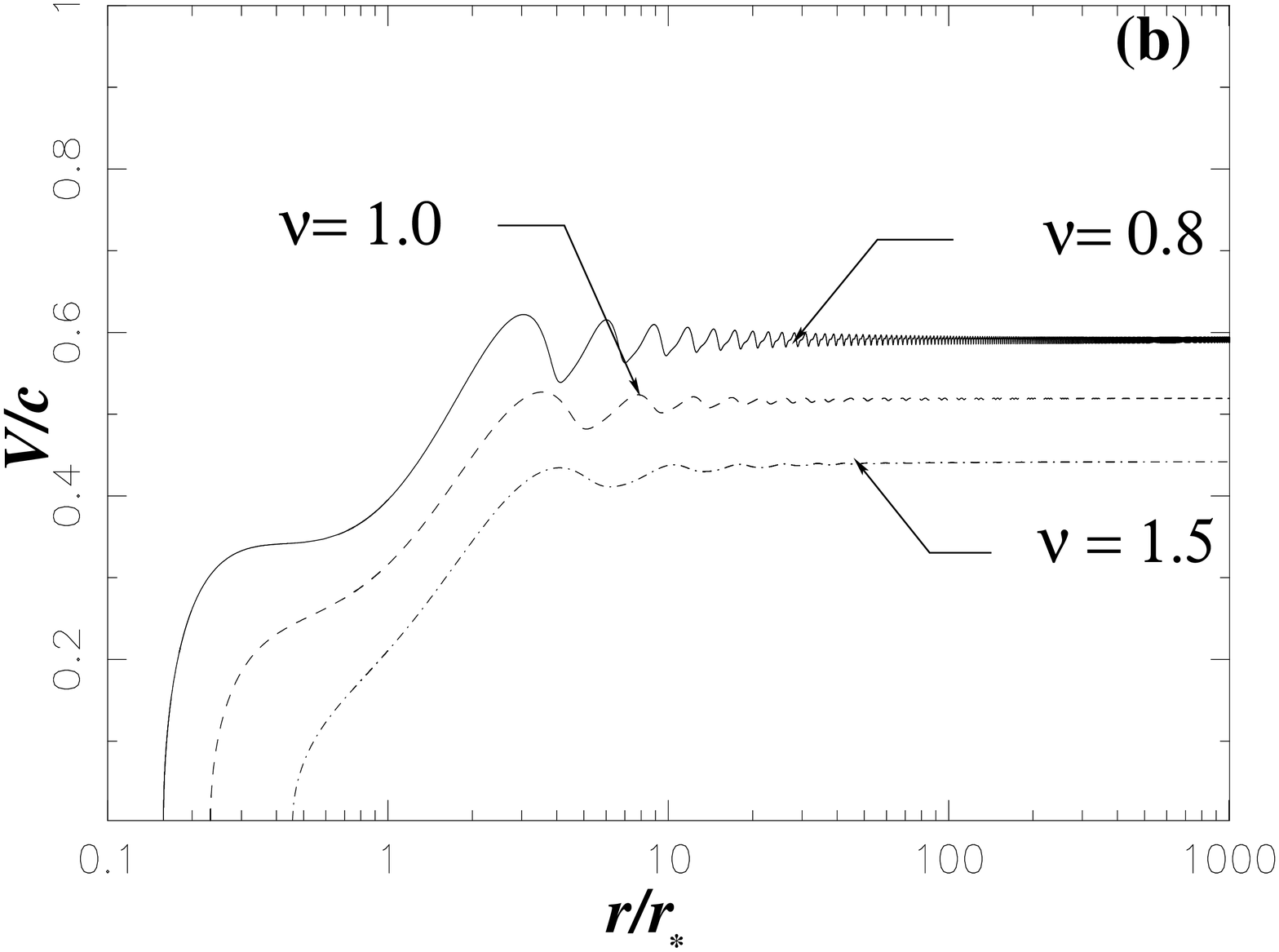}}}}
\caption{ Variation of the velocity {\it vs.} $r/r_*$ for
different values of $\mu$ in (a) and $\nu$ in (b). The other
parameters are $\nu=1.185$, $\delta = 2.113$, $\kappa = 0.5$,
$\lambda= 1.9995$ in (a), and $\mu =0.1$, $\delta = 1.0$,
$\kappa=0.5$, $\lambda=1.0$ in (b). }
\label{Jet_1_Vitesse_Diffm01}
\end{figure*}

 \section{Parametric analysis}

As as first step of the numerical analysis we have performed a
study of the effects on the solution of the free  parameters of
the model $\mu$, $\nu$, $\lambda$, $\delta$ and $\kappa$. With
respect to the classical case, the relativistic effects are ruled
by the new parameter $\mu$.

\subsection{Effect of space curvature and gravity ($\mu$, $\nu$)}

The parameters $\mu$ and $\nu$ denote the escape speed in units of
the light and Alfv\'en speed, respectively. However similar they
look, they have opposite effects on the initial acceleration and
the terminal speed. In the super-Alfv\'enic region the
acceleration is not strongly affected by different values of $\mu$
for $\mu \leq 0.1$; in fact, the effect of relativistic gravity is
negligible after $10 r_{G}$ (Fig. \ref{Jet_1_Vitesse_Diffm01}). So
the effects we are discussing now refer to the base of the flow,
in the subAlfv\'enic regime.

The parameter $\mu$, the ratio between the Schwarzschild and the
Alfv\'enic radius, representing also the escape speed at the
Alfv\`en radius in units of $c$, is related to the relativistic
effects of gravity in this model. Basically  $\mu$ controls the
extension of the corona and the acceleration of the flow in the
sub-Alfv\'enic region. Increasing $\mu$ increases also the
asymptotic velocity, as it can be seen in Fig.
\ref{Jet_1_Vitesse_Diffm01}a. A simple physical interpretation may
be given to this behaviour. When the Alfv\'enic surface approaches
the Schwarzschild surface, gravity in the sub-Alfv\'enic region,
and thus in the corona, increases. Consequently, to support
gravity the thermal energy increases too. This larger amount of
thermal energy in the corona will be transformed in turn largely
into kinetic energy along the flow. In other words, the increase
of $\mu$ implies a stronger density gradient of the flow in the
sub-Alfv\'enic region, increasing the radial pressure gradient
${\rm d} \Pi/{\rm d} R$ and leading to a stronger expansion and
acceleration.

This behaviour  can  be also understood considering that larger
values of $\mu$ imply a larger space curvature, increasing also
the expansion of the streamlines and thence the efficiency of the
acceleration, as it has been shown in the study of  the
relativistic Parker wind (see \cite{Melianietal04}).

Conversely, increasing $\nu$ decreases the asymptotic velocity as
well, since it reduces the size of the corona, keeping $\mu$
constant, that is the distance of the Alfv\'en surface to the
Black Hole (see Fig. \ref{Jet_1_Vitesse_Diffm01}b). This figure
shows that the base of the flow gets closer to the Alfv\'en radius
and farther from the Schwarzschild radius. As in the non
relativistic case the parameter $\nu^2$ is the ratio of
gravitational to kinetic energy at the Alfv\'en surface, Eq.
\ref{Defnu}. An increase of $\nu$ reduces the fluid velocity at
the Alfv\'en radius with respect to escape speed needed to get out
of the black hole's attraction. The reduction of the size of the
corona is also consistent with the reduction of the velocity. The
behaviour is as expected from the solutions in the classical
regime (ST94, STT02, STT04): the higher is the value of $\nu$ the
lower is the asymptotic velocity, although we didn't show it
explicitly, as in the present Fig. \ref{Jet_1_Vitesse_Diffm01}b.
There is also a minimum value of $\nu$ to have mass ejection,
below that value the thermal energy cannot support gravity (see
STT02).

\subsection{Effect of rotation ($\lambda$) \label{lambdaeffect}}

\begin{figure}
{\rotatebox{0}{\resizebox{8cm}{4.5cm}{\includegraphics{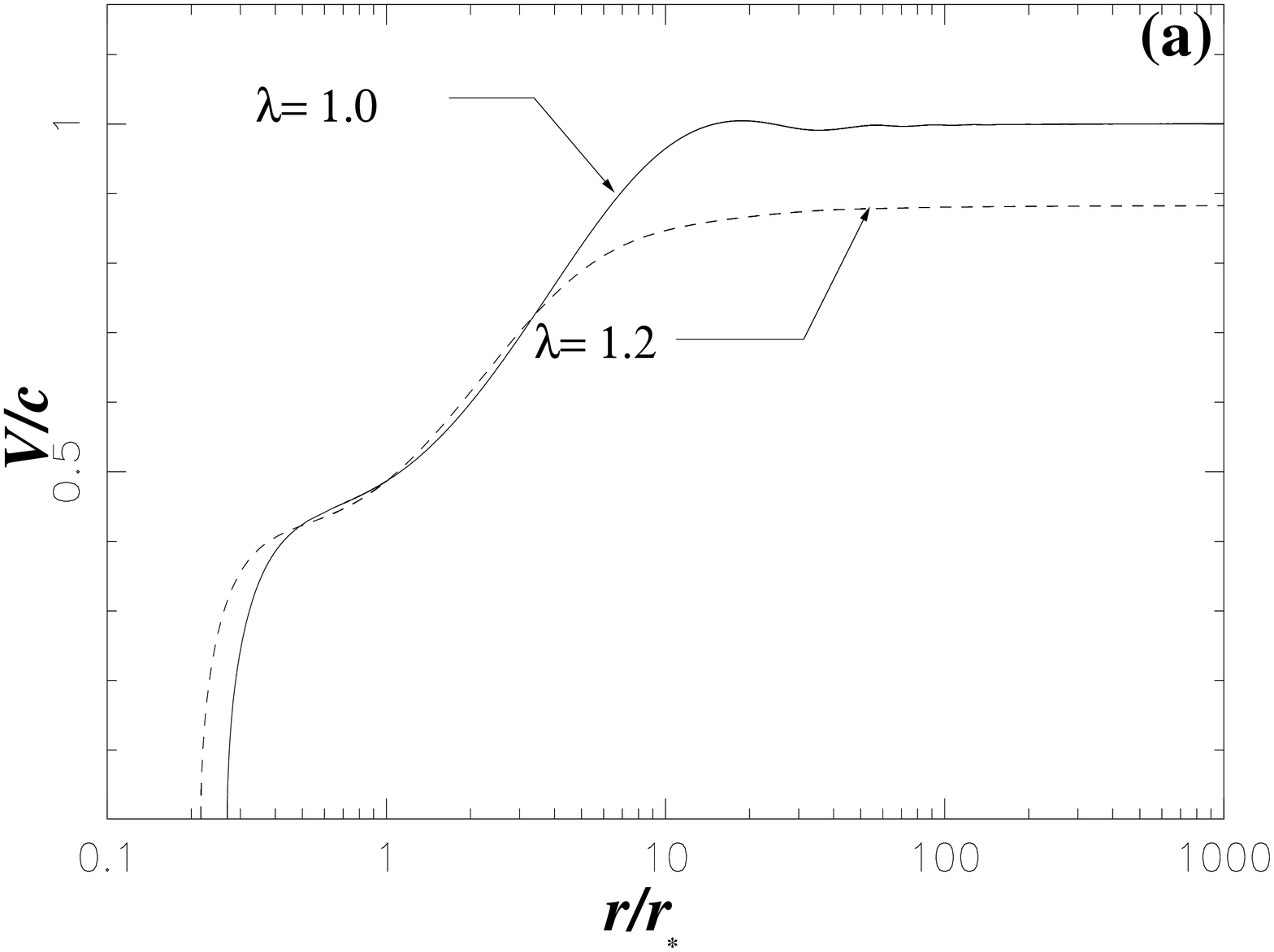}}}}

\rotatebox{0}{\resizebox{8cm}{4.5cm}{\includegraphics{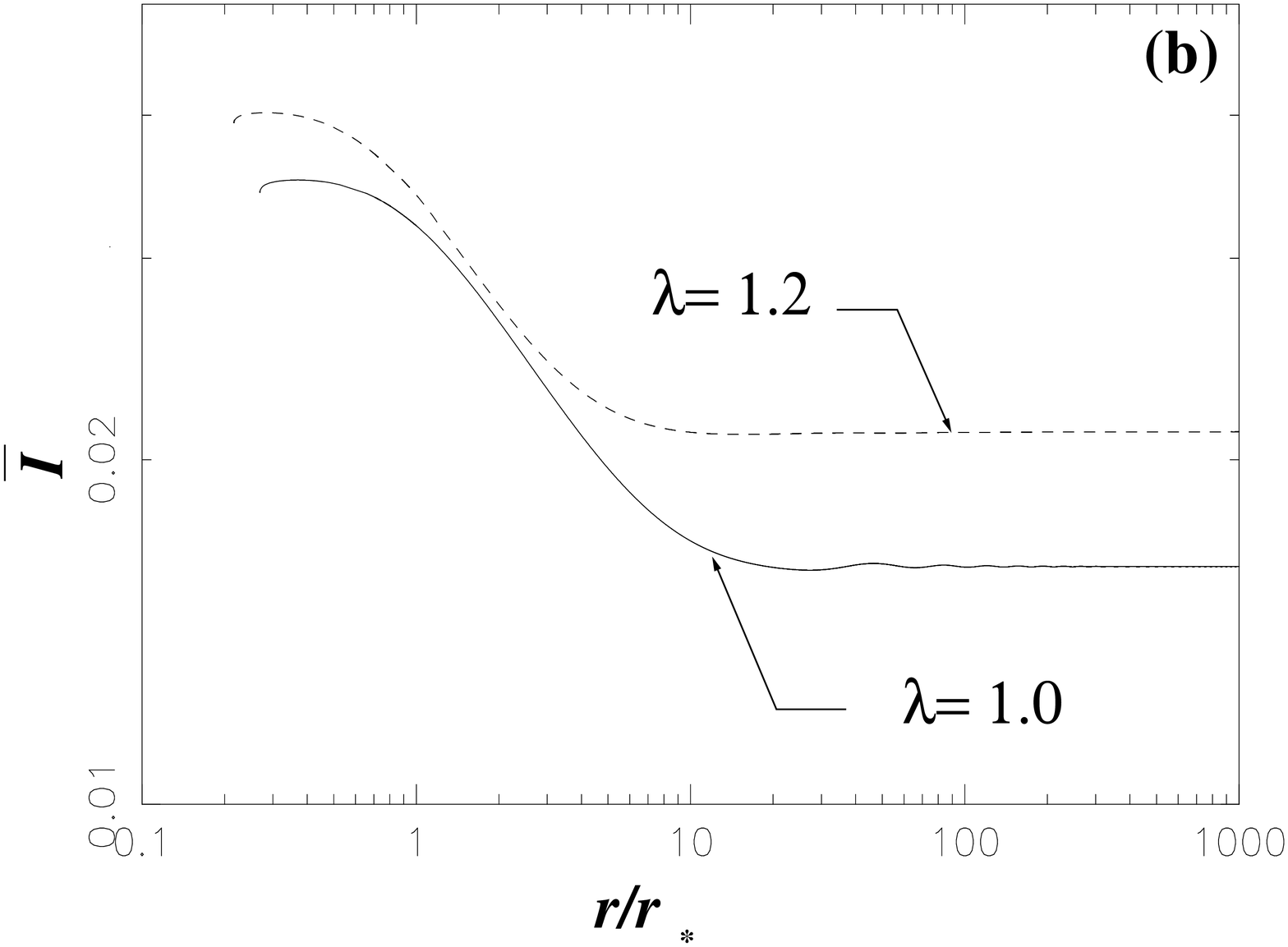}}}
\caption{Comparison between two jet solutions with $\lambda =  1.0$
($\epsilon=0.089$) and $ 1.2$ ($\epsilon=0.824$). In  (a) we plot
the velocity along the polar axis and in (b) the dimensionless electric current
density along the polar axis.
The other parameters are $\mu = 0.1$, $\nu = 0.65$, $\delta = 1.4$,
$\kappa = 0.2$ and $\Pi_{\star} = 0.75$.}
\label{Jet_1_Fig_Compare_Lambda}
\end{figure}

The parameter $\lambda$ is related to the rotation of the flow and to the
axial electric current (Fig. \ref{Jet_1_Fig_Compare_Lambda}b).
As for non relativistic outflows (ST94, STT02) it  rules the jet dynamics
through the Lorentz force, collimating asymptotically the jet via the toroidal
magnetic field, while the  centrifugal force has instead a decollimating
effect. We have checked that the behaviour of the solutions is similar to the
classical case. The increase of $\lambda$ leads to more collimated and slower
jets (Fig. \ref{Jet_1_Fig_Compare_Lambda}a). This can be understood as follows.
 Increasing $\lambda$ increases the axial current (Fig. \ref{Jet_1_Fig_Compare_Lambda}b) which increases the toroidal
magnetic pinching. In order to preserve equilibrium the flows reacts by increasing the centrifugal
force and thus its rotational speed by reducing its cross section. The reduction of the expansion
factor reduces as usual the pressure gradient and the thermal driving efficiency thus reducing the
asymptotic speed.

\begin{figure*}
{\rotatebox{0}{\resizebox{8cm}{8cm}{\includegraphics{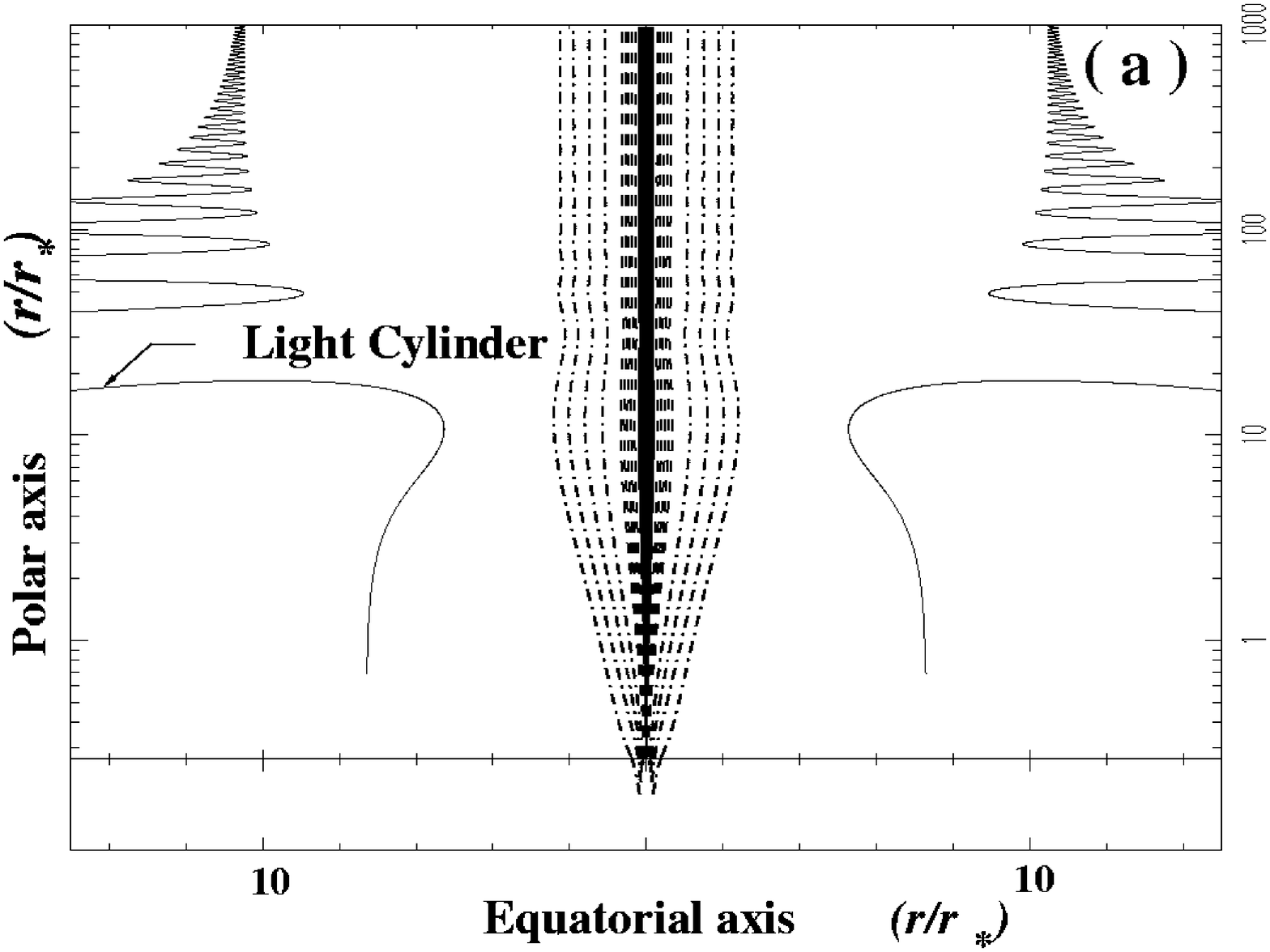}}}}
\rotatebox{0}{\resizebox{8cm}{8cm}{\includegraphics{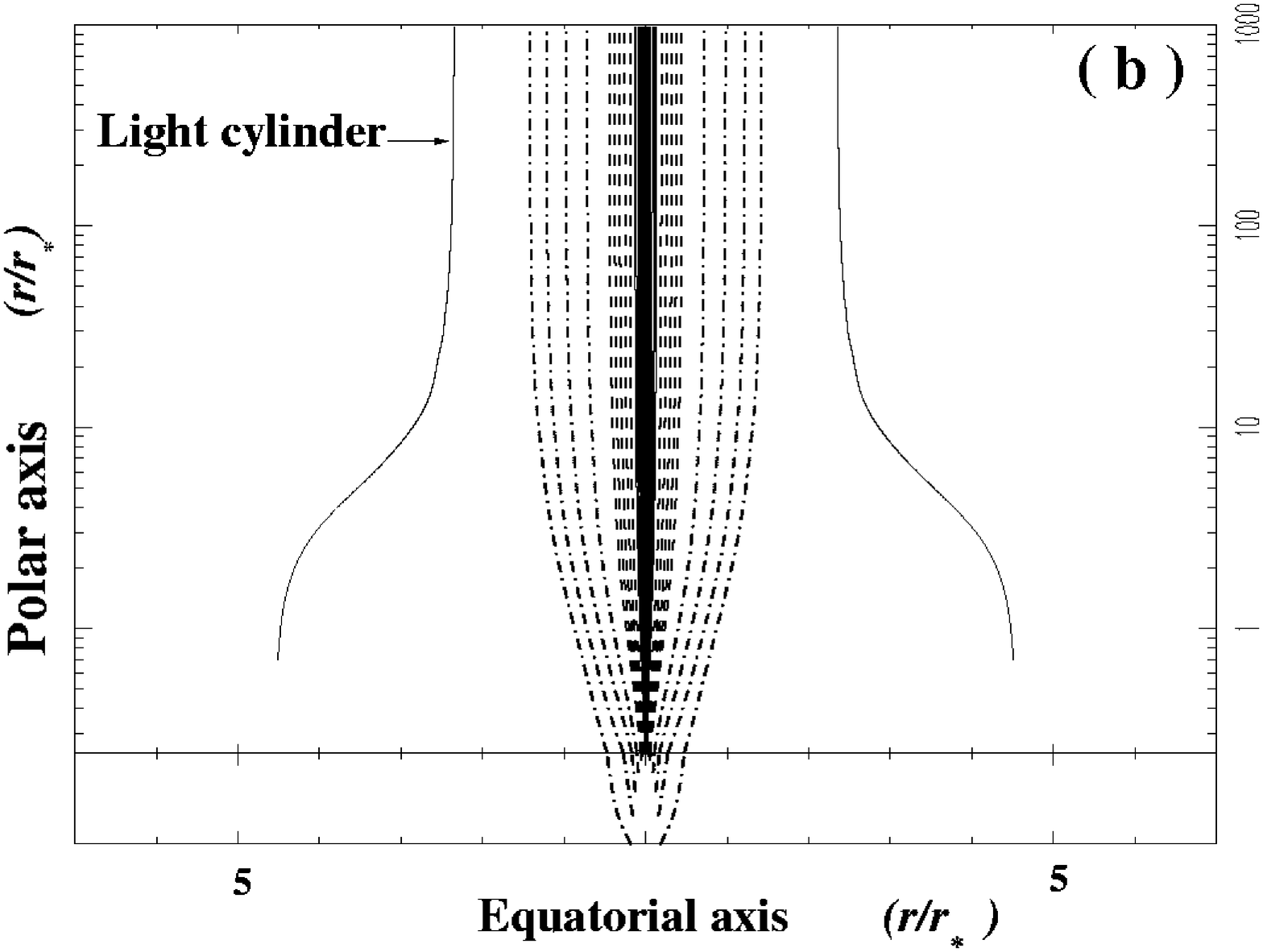}}}
\caption{
Morphology in the poloidal plane of the streamlines of the
two solutions of the previous figure and their light cylinders for
$\lambda =  1.0$ ($\epsilon=0.089$) in (a) and $ \lambda= 1.2$
($\epsilon=0.824$) in (b).
As in Fig. \ref{Jet_1_Fig_Compare_Lambda}, the other parameters are
$\mu = 0.1$, $\nu = 0.65$, $\delta = 1.4$, $\kappa = 0.2$ and
$\Pi_{\star} = 0.75$.
The light cylinders are represented by the two thick solid lines which
surround the jets, and the different regions of validity of our solutions are
shown: solid, dashed and dashed-dotted lines correspond to streamlines where
the two quantities (defined in Sec. 2.2)  $x_{\rm A}^2 G^2$ and
$(2 + \delta \alpha)\left(1 + \delta \alpha\right)^2 - 2$ are $<0.01$, $<0.1$
and $>0.1$, respectively.}
\label{Jet_1_Fig_Compare_Lambda_Morph}
\end{figure*}

In addition we must take into account that $\lambda$ is related to
the electric potential $\Phi$. Consequently it controls the
charge separation and the corresponding electric
force $\left(\rho_e  E \propto x^2 \propto \lambda^2\right)$.
This force becomes dominant where the jet rotation speed becomes
relativistic $\left(\varpi \Omega\sim c\right)$. In other words the higher is
$\lambda$ the larger the effects of the light cylinder
(Fig. \ref{Jet_1_Fig_Compare_Lambda_Morph}).

\subsection{Effect of pressure and density anisotropies ($\kappa$, $\delta$)}

The physical meaning of $\kappa$ remains the same as in non
relativistic flows. For $\kappa$ positive or negative the gas
pressure increases or decreases with colatitude, respectively.
Then in the first case the gas contributes to the thermal
confinement of the flow (underpressured jets), while in the second
to its thermal support (overpressured jets). We have limited
ourselves to present here the study of underpressured flows. The
behaviour of the relativistic solutions  with $\kappa$ is
analogous to that of  classical flows. For higher $\kappa$ both
the asymptotic velocity and jet cross sections decrease (see STT99
and STT02 for details). 

In the present model the parameter $\delta$ controls the variation of the ratio $n/w$ with
colatitude, or equivalently in the direction perpendicular to the rotational axis.
This is a relativistic generalization of the classical solutions where it
governs the transverse profile of the mass density. As a result,
the effects are similar to those found for non relativistic solutions.
A larger $\delta$ means a larger gravitational potential of
the external streamlines with respect to the axis, where the acceleration is more efficient. Then
the asymptotic velocity increases with $\delta$ (ST94, STT02).

\begin{figure*}
\rotatebox{0}{\resizebox{8cm}{7.5cm}{\includegraphics{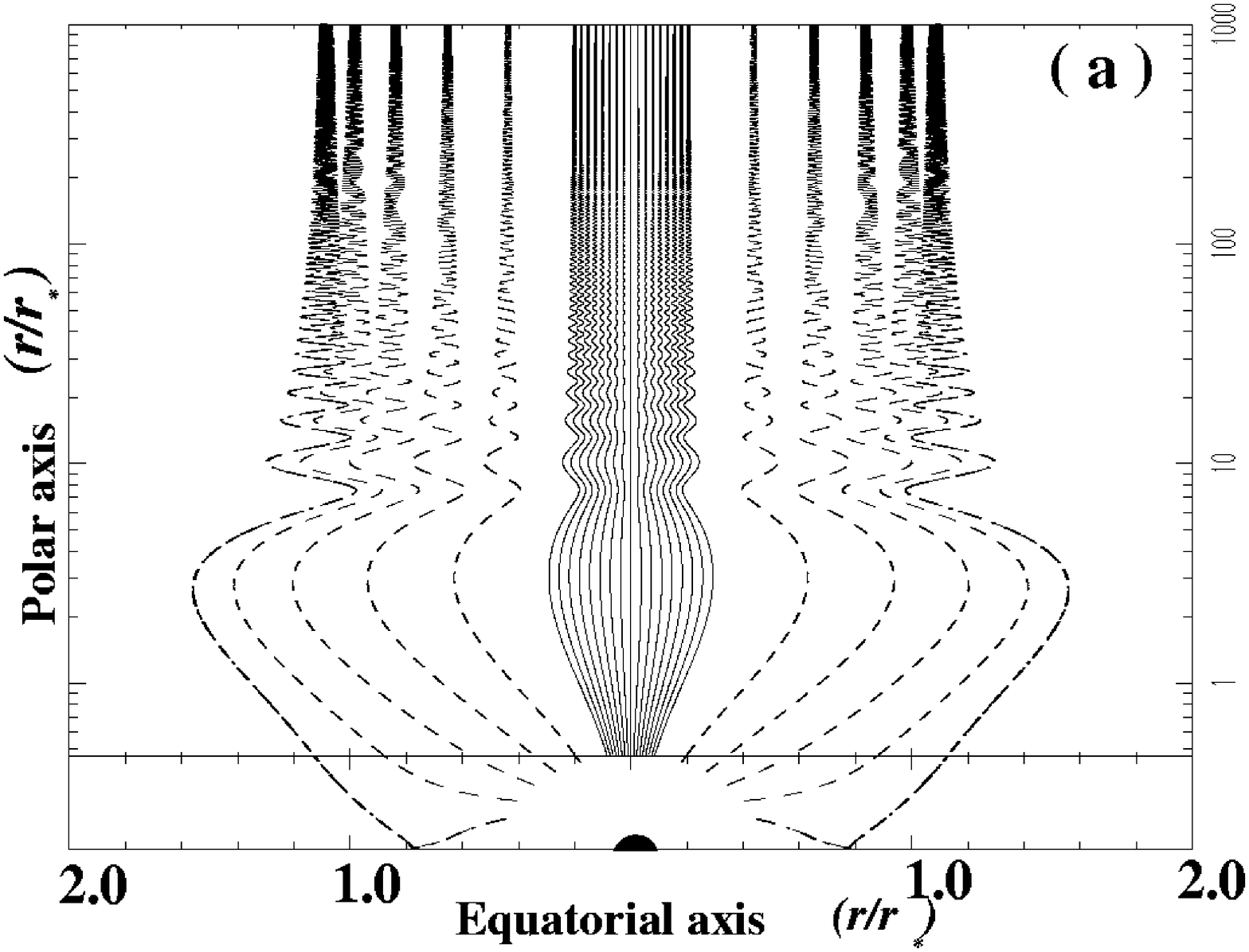}}}
{\rotatebox{0}{\resizebox{8cm}{7.5cm}{\includegraphics{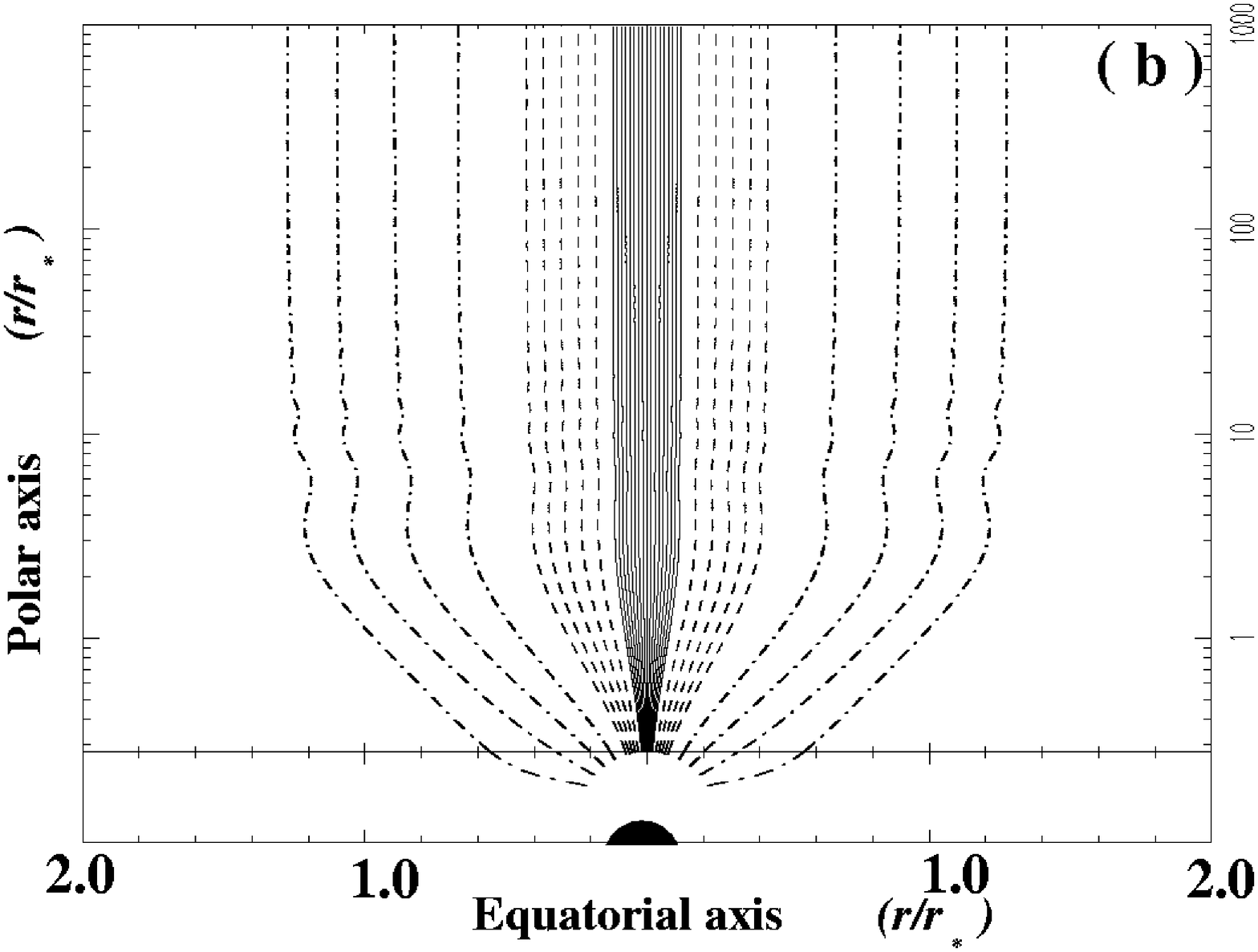}}}}
\caption{Morphology of the poloidal streamlines for two solutions
corresponding to an Inefficient Magnetic Rotator (IMR, $\epsilon=
-1.747$) in (a), and an Efficient Magnetic Rotator (EMR,
$\epsilon= 1.128$) in (b). We chose $\mu=0.1$  in both cases,
while the other parameters are $\nu=0.781$, $\kappa=0.49$,
$\delta=2.613$, $\lambda=0.880$ and $\Pi_\star=1.4$ in (a), and
$\nu=0.541$, $\kappa=0.39$ , $\delta=3.253$, $\lambda=1.401$ and
$\Pi_\star=1.26$ in (b). The various regions of validity of  our
solutions are shown as in Fig.
\ref{Jet_1_Fig_Compare_Lambda_Morph}.} \label{Jet_1_FigMorp01}
\end{figure*}

\begin{figure*}
{\rotatebox{0}{\resizebox{8cm}{4.5cm}{\includegraphics{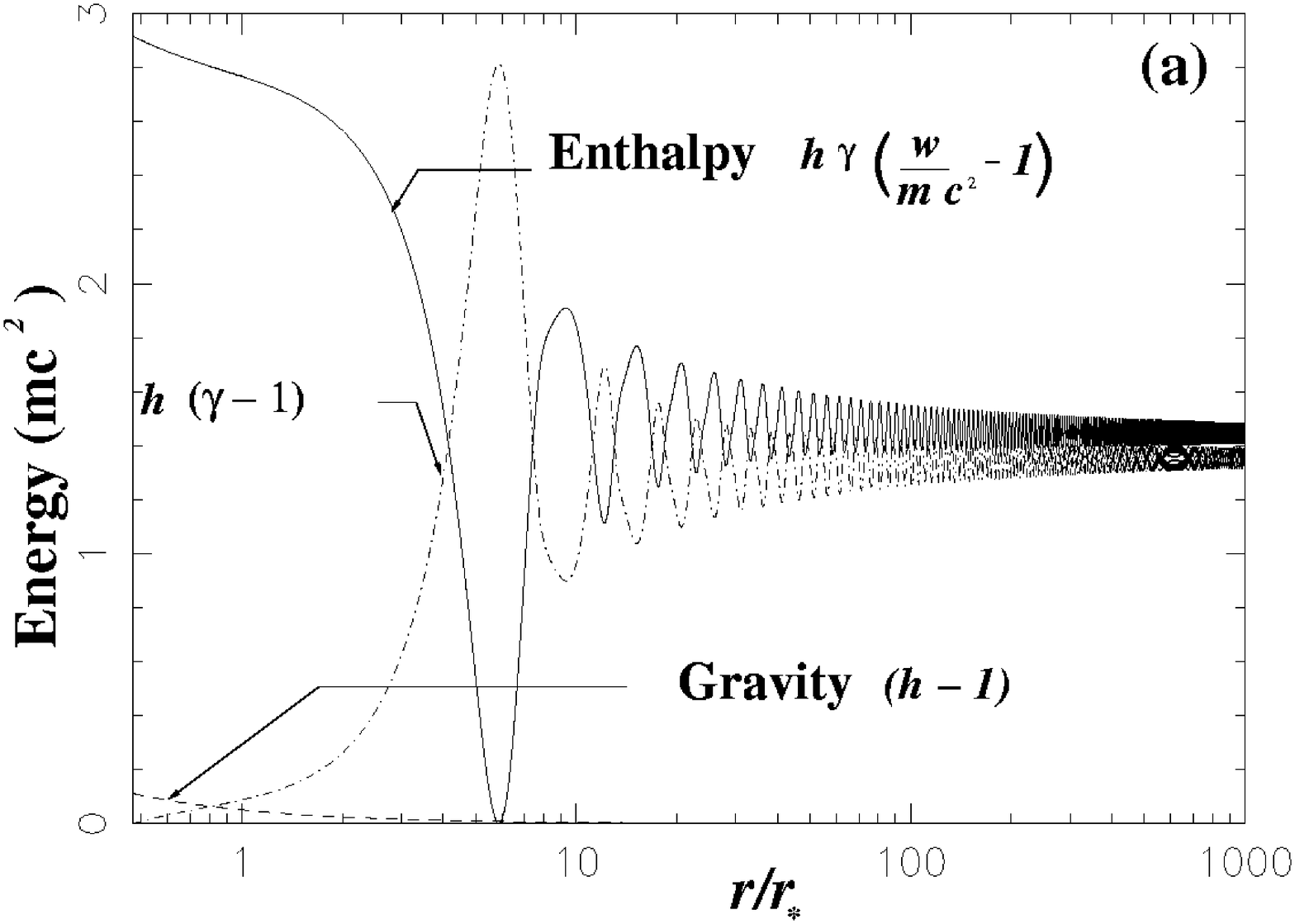}}}}
\rotatebox{0}{\resizebox{8cm}{4.5cm}{\includegraphics{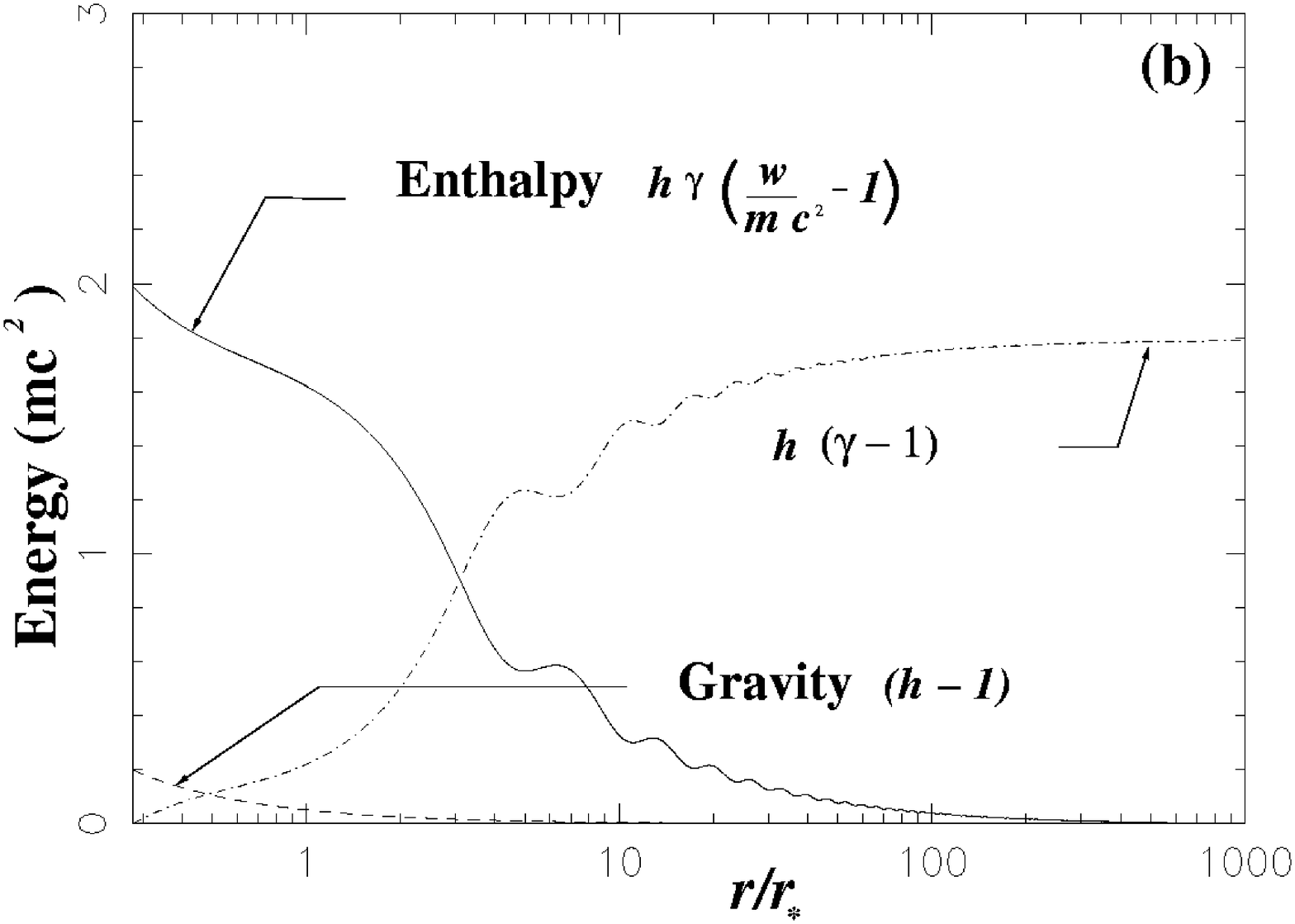}}}
\caption{Variation of the energy flux normalized to the mass
energy, along the external streamline for the IMR solution in (a)
and the EMR solution in  (b) of the  previous figure. The Poynting
energy is $\leq$ 1\% of the enthalpy at the base of the flow and
is not plotted. The parameters are the same as in Fig.
\ref{Jet_1_FigMorp01}.} \label{Jet_1_Figenergy01}
\end{figure*}

\begin{figure*}
{\rotatebox{0}{\resizebox{8cm}{4.5cm}{\includegraphics{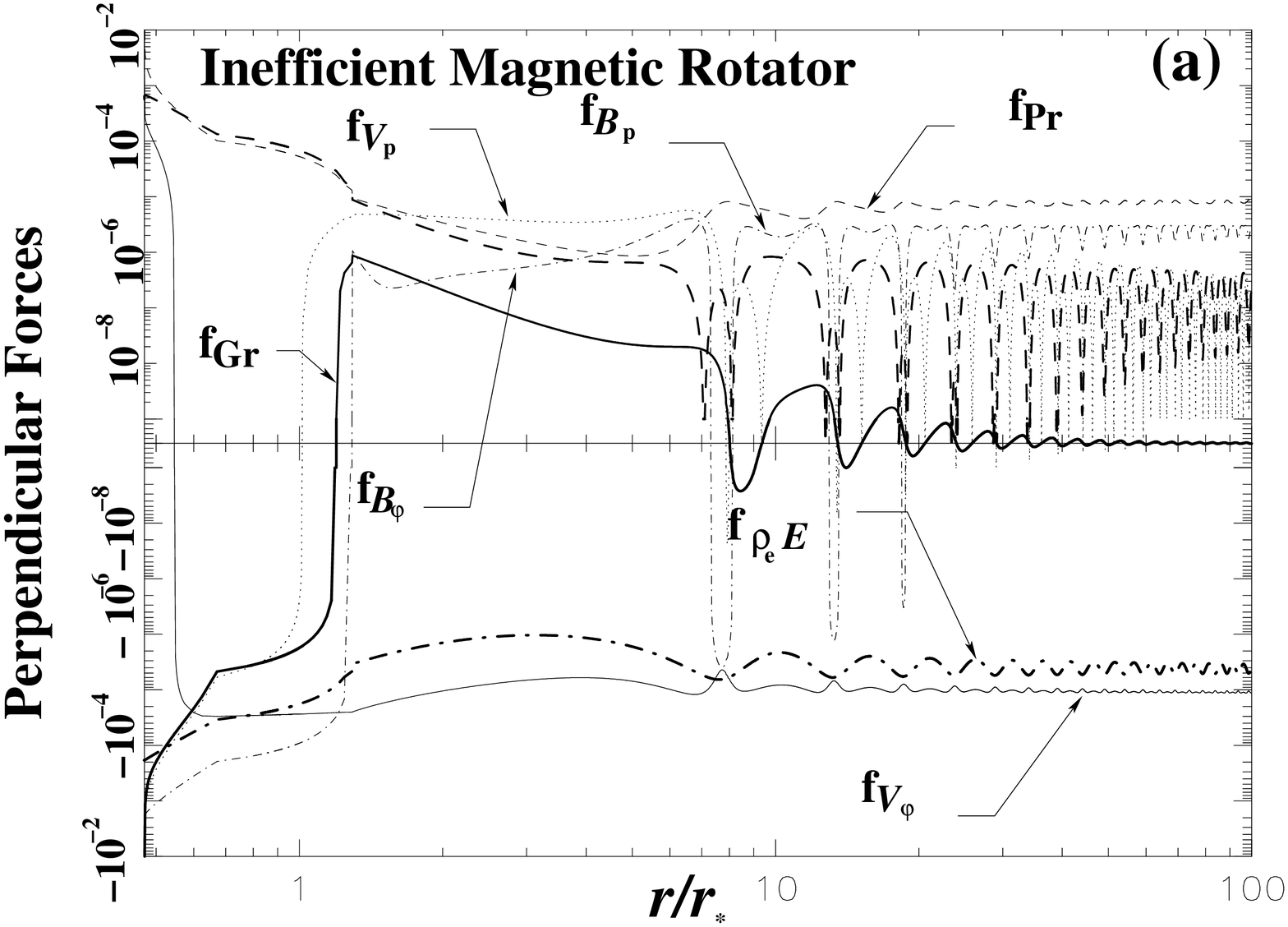}}}}
\rotatebox{0}{\resizebox{8cm}{4.5cm}{\includegraphics{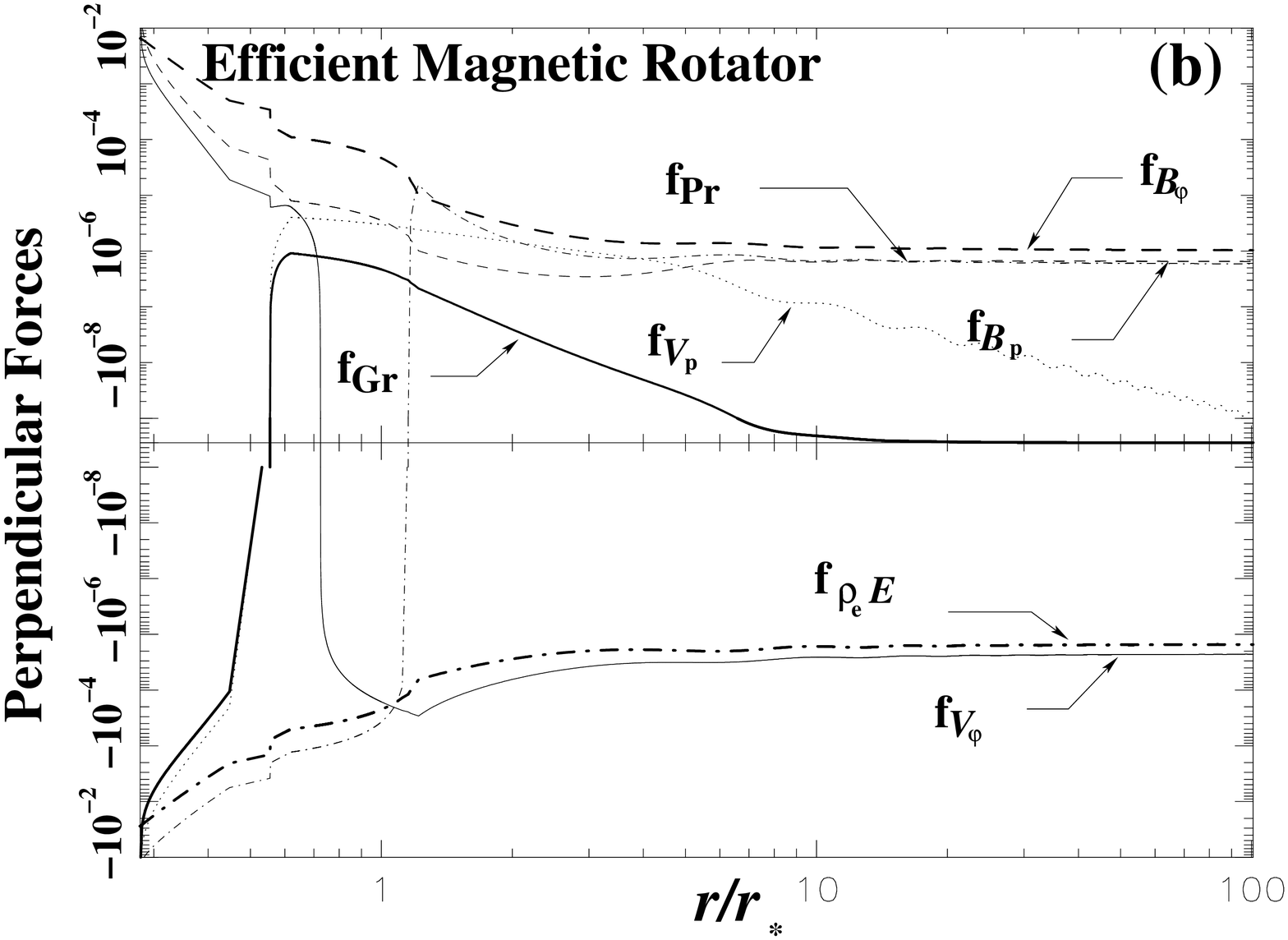}}}

\caption{{ Plot of the transverse forces  for the relativistic IMR
(a) and EMR (b).} Forces are normalized by their maximum value, usually reached
at the base of the flow. The parameters are the same as in Fig.
\ref{Jet_1_FigMorp01}. 
\label{Jet_1_FigForce_transverse_Rel}}
\end{figure*}

\section{Jet dynamics, acceleration and collimation}

We will address now the question of the process of acceleration
and collimation of the jet in the case of EMRs and IMRs. Keeping
fixed $\mu = 0.1$ we have displayed the results for an IMR
solution with $\nu = 0.781$, $\delta = 2.613$, $\kappa = 0.490$,
$\lambda = 0.880$ and $\Pi_\star=1.4$ ($\epsilon = -1.747$) 
in Fig. \ref{Jet_1_FigMorp01}a (see also Figs.
\ref{Jet_1_Figenergy01}a and
\ref{Jet_1_FigForce_transverse_Rel}a), and an EMR with $\nu =
0.541$, $\delta = 3.253$, $\kappa = 0.39$, $\lambda = 1.401$ and
$\Pi_\star=1.26$ ($\epsilon = 1.128$) in Fig.
\ref{Jet_1_FigMorp01}b (see also Figs. \ref{Jet_1_Figenergy01}b
and \ref{Jet_1_FigForce_transverse_Rel}b).

We see that the shape of the streamlines in the two cases (Figs.
\ref{Jet_1_FigMorp01}) looks similar to the corresponding non
relativistic case (STT02, see later Figs.
\ref{Jet_1_Fig_Compare_morpholgie_classique_Relativiste}). IMRs
show a fast expansion in an intermediate region, while far from
the base the collimated streamlines show strong oscillations. EMRs
show conversely a continuous  expansion with relatively mild
oscillations,  or even no oscillations at all when pressures are
lower. We also display for the two solutions the energies along a
given streamline (Fig. \ref{Jet_1_Figenergy01}) and the forces
perpendicular to the flow (Fig.
\ref{Jet_1_FigForce_transverse_Rel}).

\subsection{Acceleration}

By construction of this model, the wind is basically
thermally driven. At the lowest order we have ${\cal E} \approx h_{} \gamma w$,
while the first order term corresponds to the Poynting flux which remains 
however of
the same order than the thermal terms in the transverse direction.
This supposes that there is a high temperature corona around the black hole 
as proposed
by Chakrabarti (1989) and Das (1999, 2000). The latest has shown that the 
stronger is the thermal
energy, the more stable is the corona.

We can study the acceleration of the jet analysing the
contribution of the different energies and their conversion from
one to the other along the streamlines. The dominating energy at
the base of the outflow is the enthalpy. Part of it is used to
balance gravity and the remaining part is converted into kinetic
energy in the region of expansion of the jet and stops when the
streamlines collimate (compare Figs. \ref{Jet_1_FigMorp01} and
\ref{Jet_1_Figenergy01}). In fact, during the expansion of the
jet, the plasma density decreases which also induces a decrease of
the enthalpy. In turn, it creates a strong pressure gradient
$\left(\nabla P = n \nabla w\right)$ that accelerates the jet. For
the parameters we have chosen, we see that the EMR solution has a
larger expansion factor than the IMR one (Figs.
\ref{Jet_1_FigMorp01}), because of the thermal driving. This is
correlated to the larger Lorentz factor of the EMR solution
($\gamma\sim 2.8$) as compared to the IMR one ($\gamma\sim 2.4$).
This result is not related to the nature of the magnetic rotator. For 
other parameters, we would get different results but the
asymptotic Lorentz factor always increases with the increase of
the expansion factor because of the thermal driving. We verified
that the Poynting flux in the two solutions is negligible,
representing at maximum only $1\%$ of  the enthalpy at the basis
of the flow.

The IMR solution  undergoes a strong expansion in this region
until a distance of $100 r_{\star}$ and then recollimates and
consequently decelerates because of the compression. On the other
hand, the EMR solution  collimates already at a distance of
approximately $50 r_{\star}$ and accelerates all the  way
downwind. In other words, the nature of the collimation affects
obviously the velocity profile.

\subsection{Collimation}

The collimation of the flow is controlled by
different types of  forces that depend on the morphology of the
streamlines in the jet. We have plotted for the two solutions the
forces perpendicular to the streamlines
in  Fig. \ref{Jet_1_FigForce_transverse_Rel}.
Asymptotically the centrifugal force is the dominant term which
supports the wind  against either the magnetic confinement in EMR
or the pressure gradient in IMR.

The behaviour of the other forces depends on the shape of the streamlines
(Fig. \ref{Jet_1_FigForce_transverse_Rel}) and they play a relevant role
in the intermediate region before collimation is fully achieved.
In particular, the stress tensor from  the poloidal magnetic field
and the gravity favor deviation from radial expansion while the
thermal pressure
gradient initially tends to maintain the radial expansion. In the region of
formation of the jet, the strong gravity along the polar axis generates a
strong pressure gradient.
As density and pressure increase with colatitude ($\delta, \kappa >0$), it also
generates a total force  $f_{\nabla_{\bot} P} + f_{\nabla_{\bot}\ln h_{}}
\propto \nu^2 (\delta-\kappa)$ which further out in the jet  provides the
thermal confinement.

In an IMR, neither the pressure gradient nor the pinching force
from the toroidal magnetic field can brake the expansion of the
flow, and the recollimation occurs where the curvature of the
poloidal streamlines becomes relevant. In the asymptotic region,
the collimation is mainly provided by the transverse pressure
gradient, $\kappa \Pi$, which balances the centrifugal force. The
pressure confined jets undergo strong oscillations similarly to
the non relativistic case (STT94, STT99).

In an EMR, conversely, the pinching force of the
toroidal magnetic field provides collimation all along the flow.
The pressure gradient may help this collimation as for the present
solution or be completely negligible for other sets of parameters.
The magnetic pinching force is balanced asymptotically mainly by the
centrifugal force which tends to decollimate the jet.
We must notice also that, as expected in the relativistic case, the electric
force is always positive and its effect in decollimating the jet may be
comparable with the centrifugal force, differently from IMRs
(see Fig  \ref{Jet_1_FigForce_transverse_Rel}).

\begin{figure*}
{\rotatebox{0}{\resizebox{8cm}{7.5cm}{\includegraphics{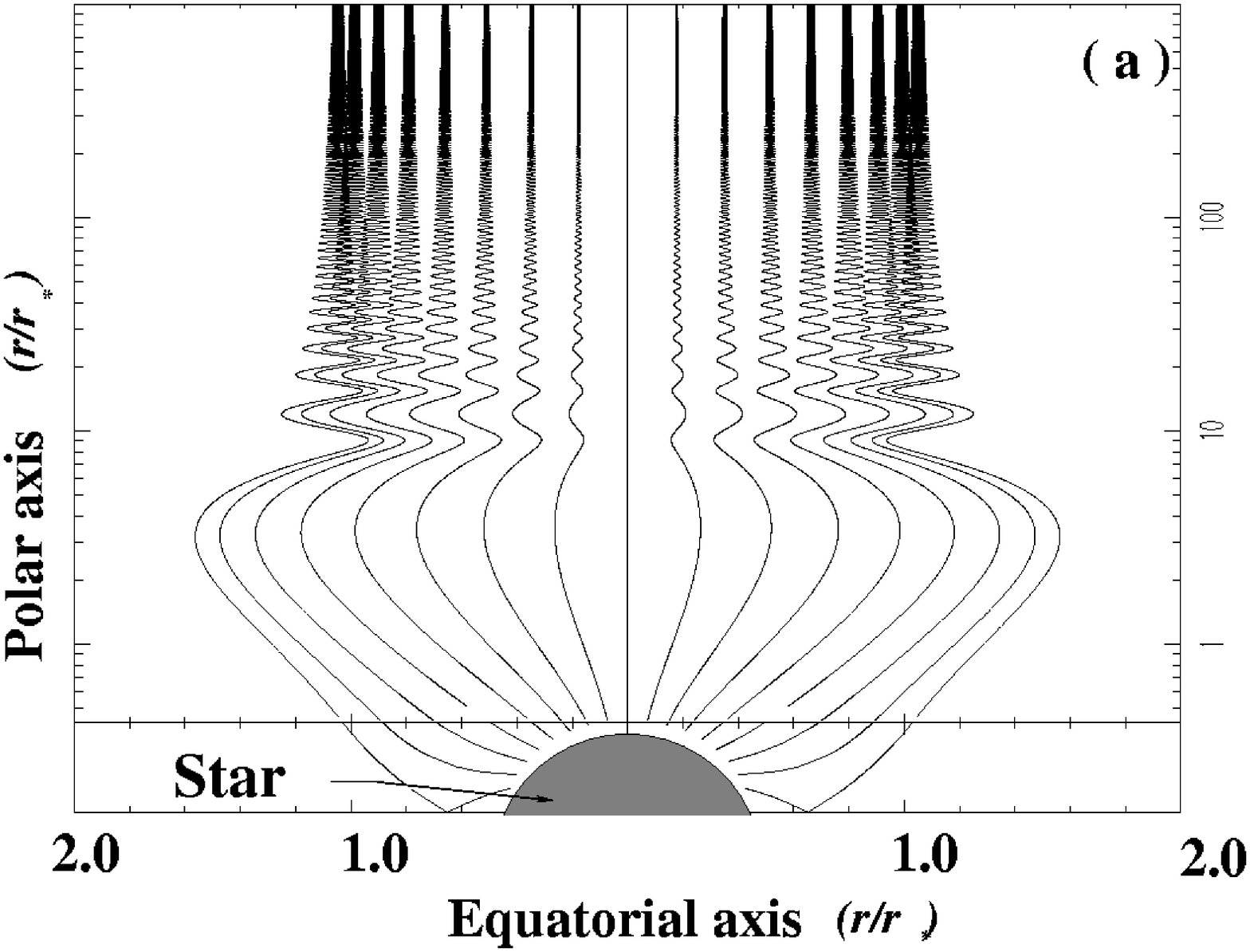}}}}
\rotatebox{0}{\resizebox{8cm}{7.5cm}{\includegraphics{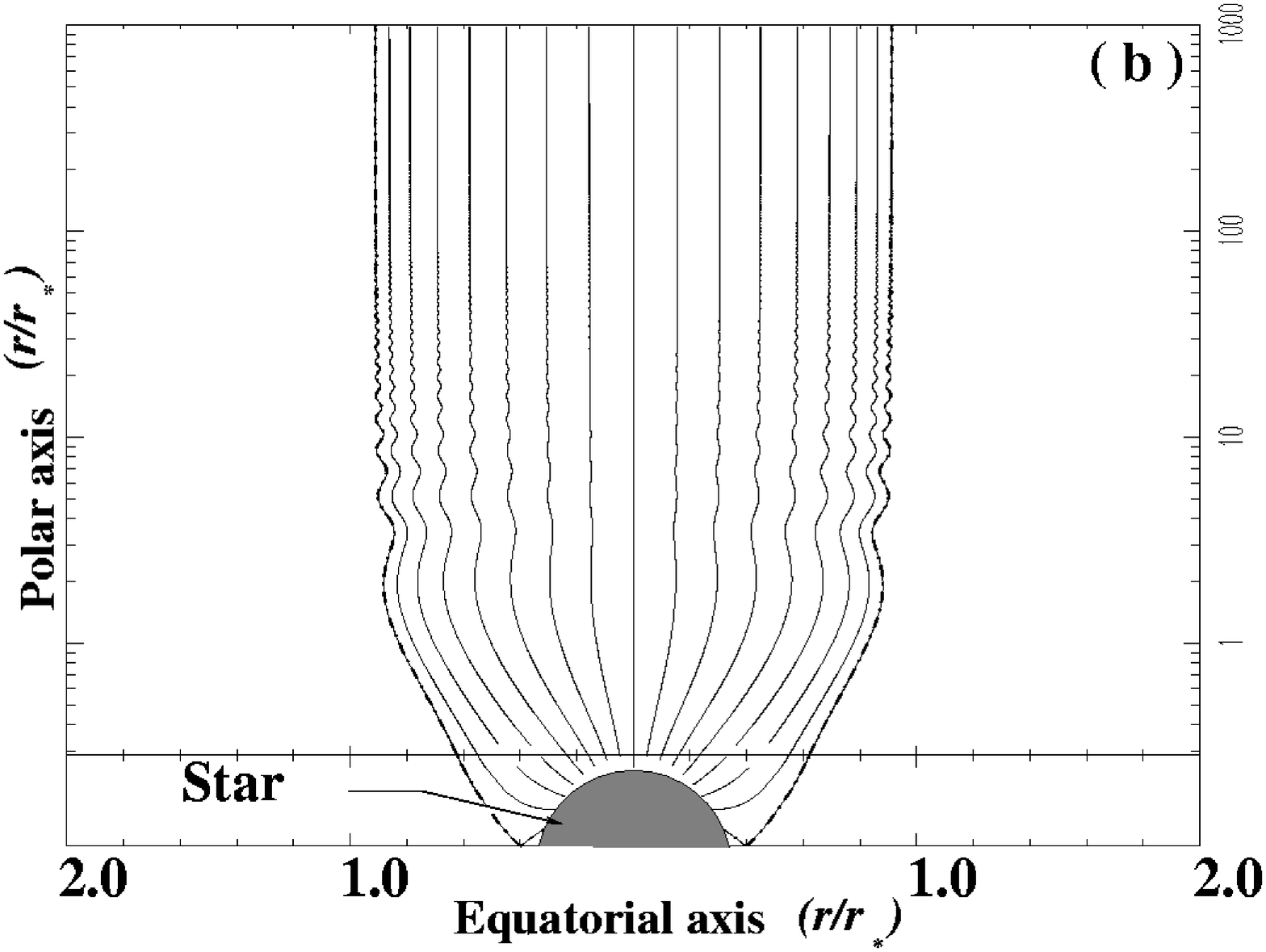}}}\caption{ Morphology of the poloidal streamlines for non
relativistic IMR (a) and EMR (b): the parameters are the same as in Fig. 4. with $\mu = 10^{-5}$.}
\label{Jet_1_Fig_Compare_morpholgie_classique_Relativiste}
\end{figure*}
\begin{figure*}
{\rotatebox{0}{\resizebox{8cm}{6cm}
{\includegraphics{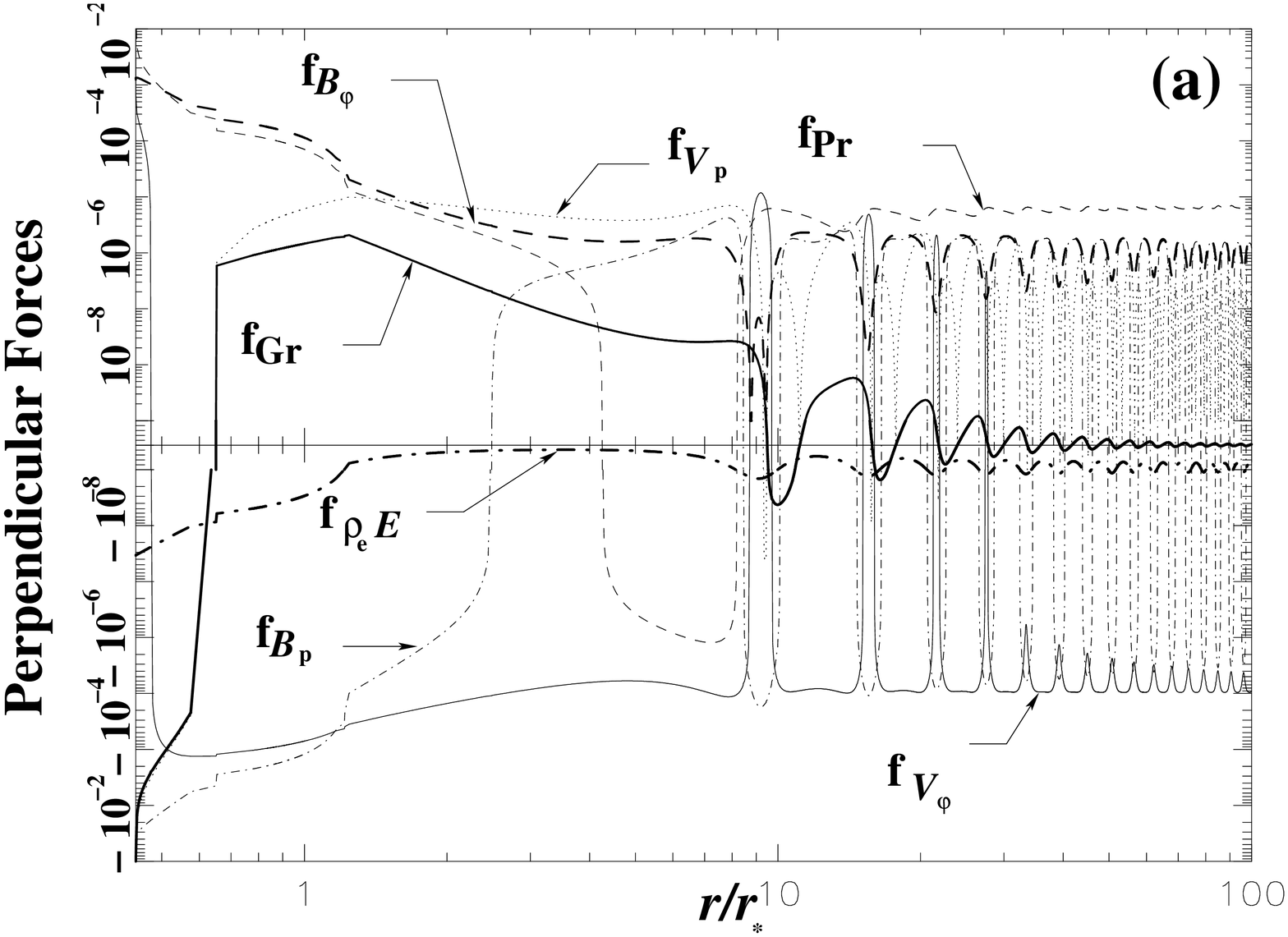}}}}
\rotatebox{0}{\resizebox{8cm}{6cm}
{\includegraphics{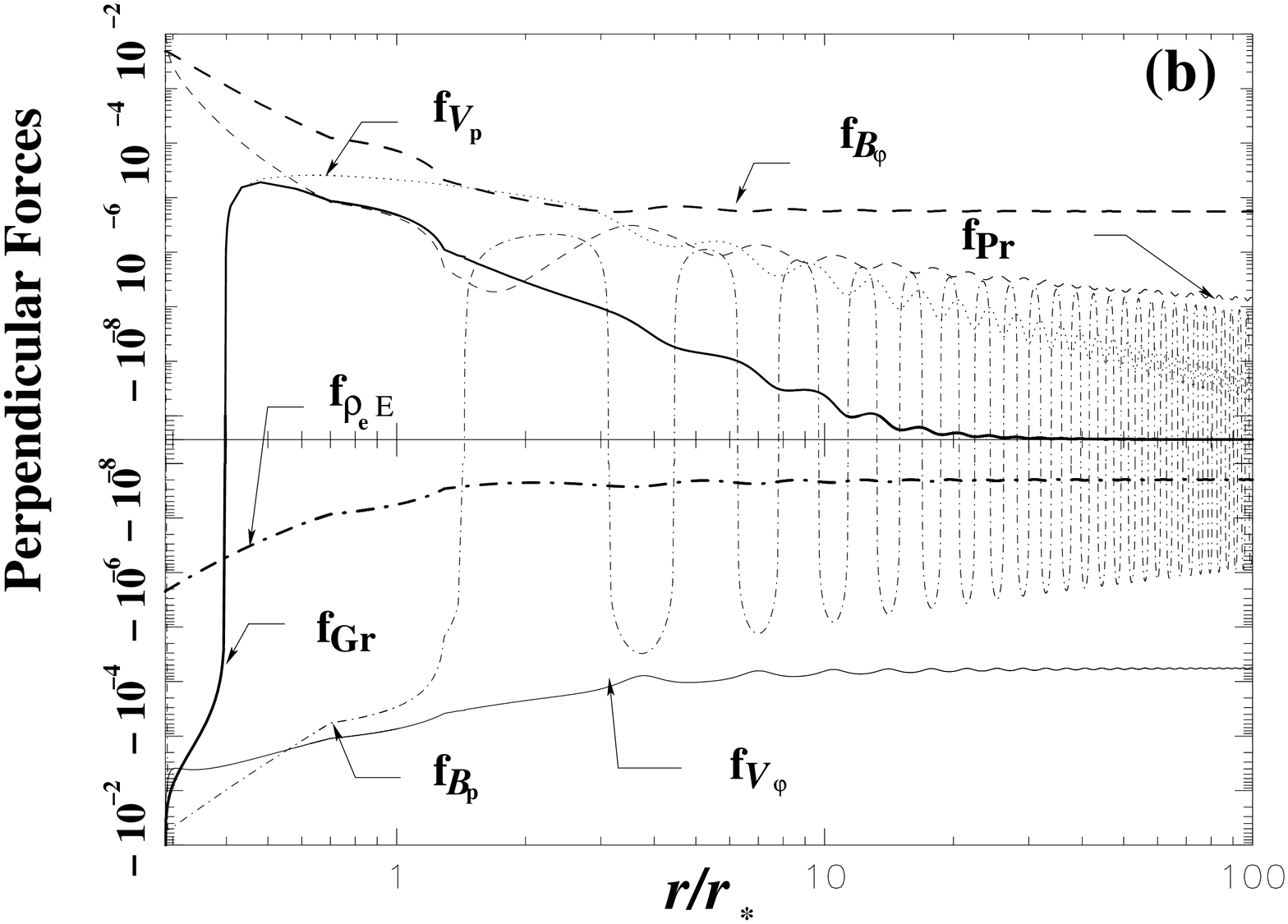}}}
\caption{Plot of the transverse forces  for the non relativistic IMR
(a) and EMR (b). We assumed
$\mu = 10^{-5}$ while the other parameters are identical to the corresponding
relativistic solutions displayed in Figs. \ref{Jet_1_FigMorp01}.
\label{Jet_1_FigCompare_Force__Classique}}
\end{figure*}

\begin{figure}
{\rotatebox{0}{\resizebox{8cm}{4.5cm}{\includegraphics{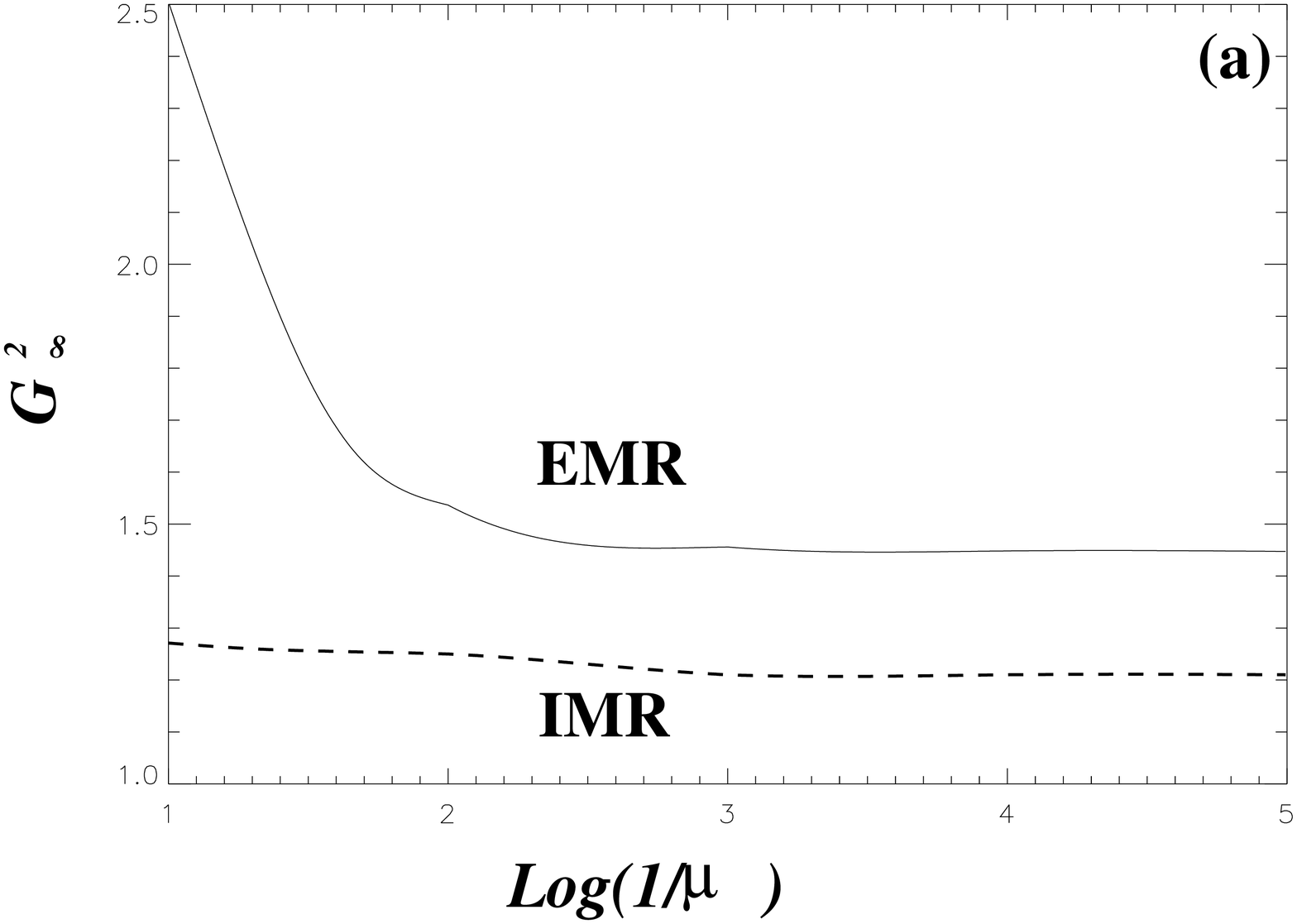}}}}

\rotatebox{0}{\resizebox{8cm}{4.5cm}{\includegraphics{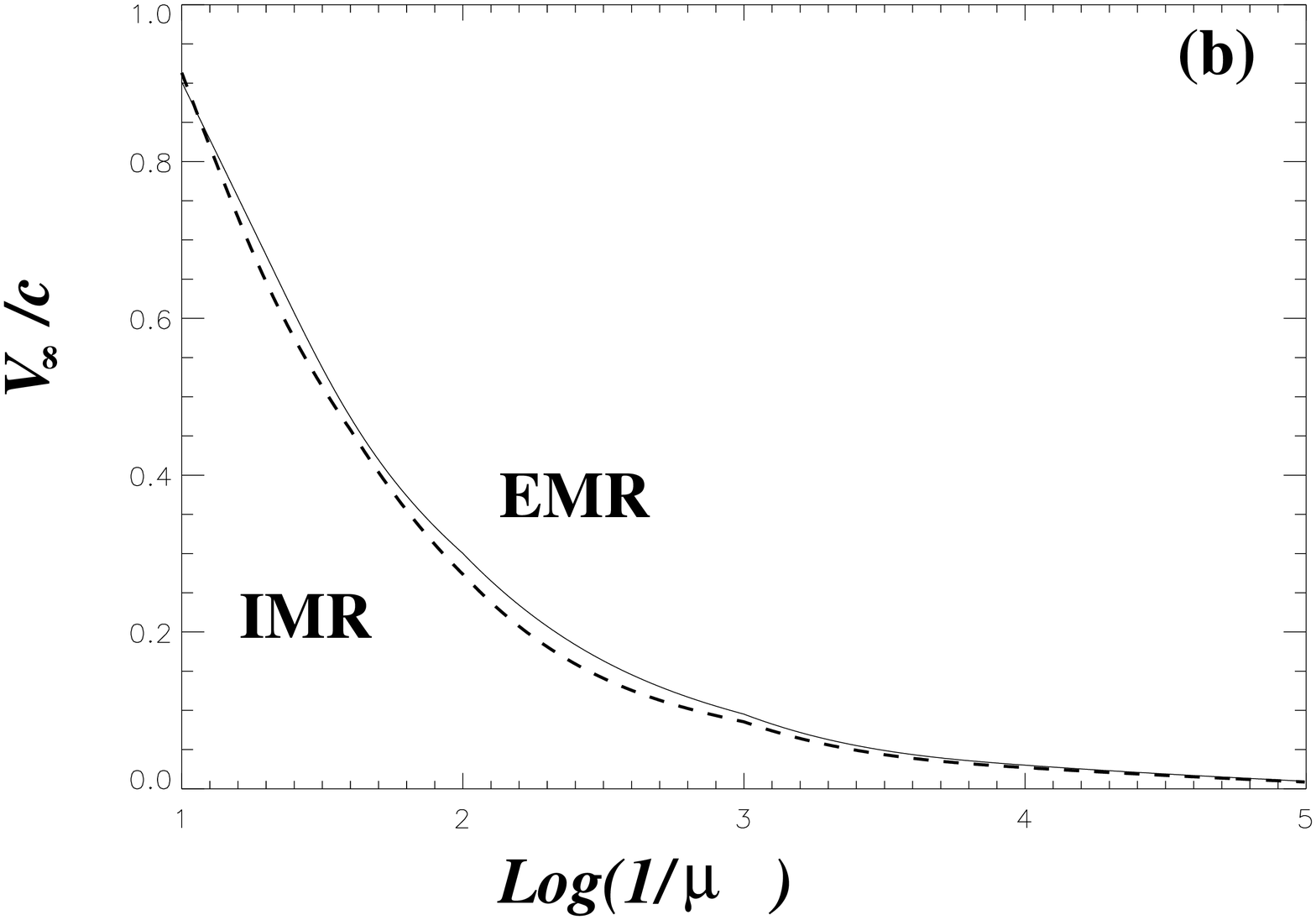}}}
\caption{Plot of the asymptotic dimensionless cross section of the
jet $G^2_{\infty}$  in (a),  and  the asymptotic
velocity in  (b) as a function of the parameter  $\mu$ on a  logarithmic scale.
The other parameters are those given for the previous relativistic
IMR  and EMR solutions.
\label{Jet_1_Application_M2}}
\end{figure}

\section{Relativistic vs. non relativistic outflows}

We analyse here in more detail the main differences between the
relativistic and non relativistic  solutions already discussed
in previous Sect. 6. By increasing
the value of $\mu$ we increase the depth of the potential well. In our calculations we have assumed
$\mu =0.1$ for the relativistic solutions and this can be justified as follows.

We supposed that the Alfv\'en surface is roughly at a distance of
$10r_{o}$ from the central object. This typical distance is usually chosen
because it corresponds to the case where the wind
carries away all the accreted angular momentum, provided about 10\% of
the accreted mass goes to the jet (\cite{Livio99}).  This is of course arbitrary but
allows us to compare our solutions to other models.
In the case of young stellar jets, the star has a mass of the order of
$\sim 1 M_{\odot}$. The corresponding Schwarzschild radius is approximately
$r_{G} \approx 3$ km. Therefore, space curvature
at the Alfv\'en surface corresponds to a value of $\mu\approx 10^{-5}$. Conversely, for
AGN jets, with a central super massive black hole of $\sim 10^9 M_\odot$, we have
$r_{G}\approx 10^{4} R_{\odot}$ which corresponds to the value we have chosen $\mu\approx 0.1$.

We have drawn the corresponding morphologies of two non relativistic solutions associated with an
IMR and an EMR in Figs.
\ref{Jet_1_Fig_Compare_morpholgie_classique_Relativiste}, keeping the other 
parameters as in Fig.
\ref{Jet_1_FigMorp01}. We also plotted the corresponding transverse forces 
for the non relativistic
solutions in Figs. \ref{Jet_1_FigCompare_Force__Classique}.

\subsection{Jets from Inefficient Magnetic Rotators (IMR)}

Let's first turn our attention to the solution from an IMR. The
morphologies of  classical and  relativistic jets show indeed
small  differences. The relativistic jet, though, undergoes an
expansion slightly more important in the intermediate region (Fig.
\ref{Jet_1_FigMorp01}a) than in the classical case (Fig.
\ref{Jet_1_Fig_Compare_morpholgie_classique_Relativiste}a). This
expansion induces a slight relative increase of the curvature
forces (inertial and magnetic) compared to other forces. In the
asymptotic region, the relativistic jet recollimation is
comparable to the non relativistic one (Fig.
\ref{Jet_1_Application_M2}a).

Note also that in the asymptotic region, the relativistic jet pinching by the
toroidal magnetic field is almost null, while in the non relativistic  
solution this
force is of the order of the pressure gradient (Figs. 
\ref{Jet_1_FigForce_transverse_Rel}a and
\ref{Jet_1_FigCompare_Force__Classique}a).
This behaviour is a consequence of the decrease of the
collimation efficiency in relativistic jets.

Last, we know that in the relativistic solutions, there is
an extra electric force, $\rho_e  E$,  due to the non negligible
charge separation, which also decollimates. In IMRs,
its influence remains however weak because of the low
magnetic field.
\begin{figure}[h]
\begin{center}
{\rotatebox{0}{\resizebox{8cm}{4.5cm}{\includegraphics{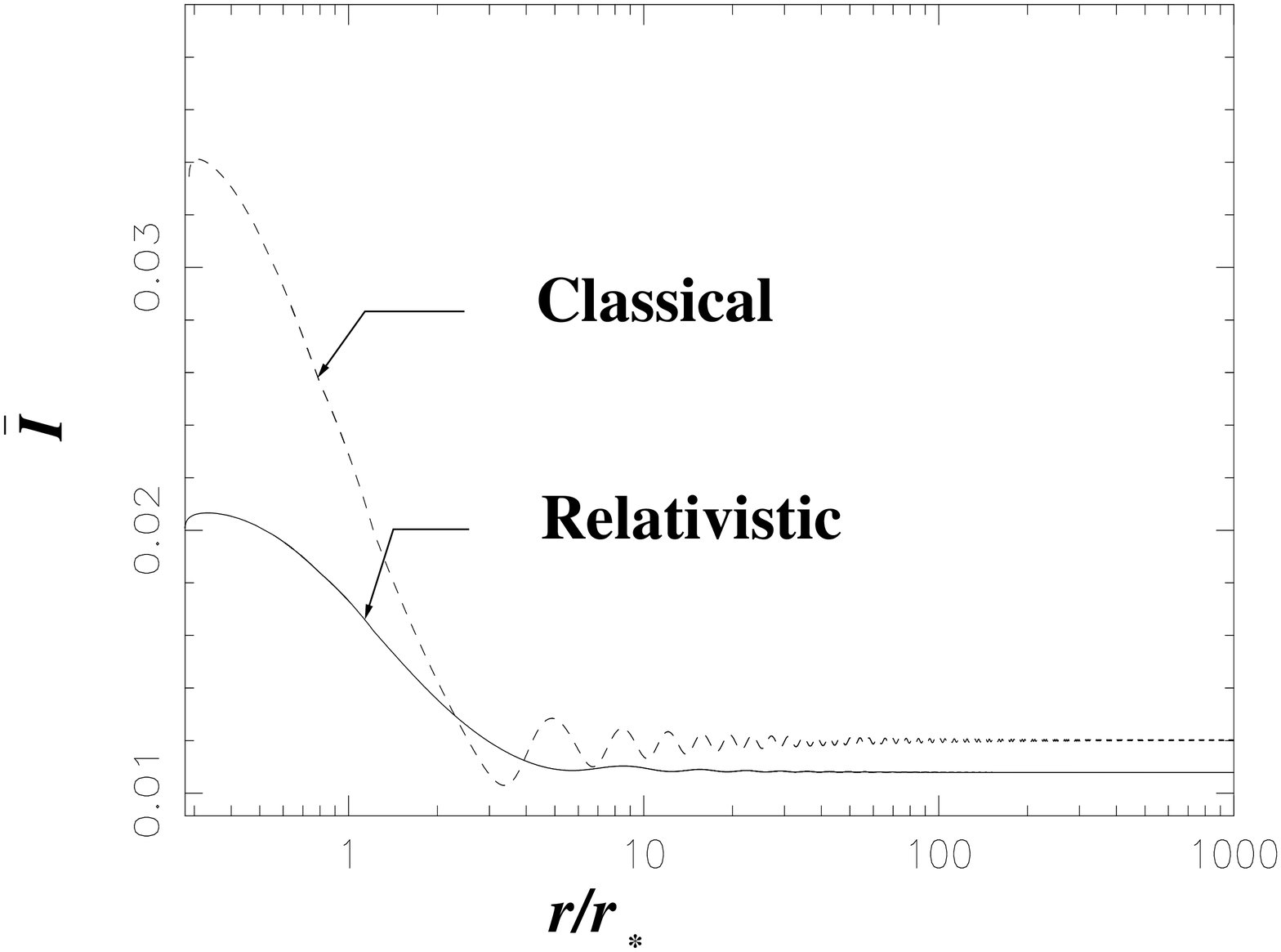}}}}
\caption{ Density of the electric current normalized $\bar{I}_{z}$,
in the classical and the  relativistic jets for EMR shown in the previous 
figures.
\label{Jet_1_Current_electrique_compare}}
 \end{center}
\end{figure}
\subsection{Jets from  Efficient Magnetic Rotators (EMR)}

In the jet solution associated with an  EMR, the solution
is effectively magnetically collimated because of the toroidal
magnetic field pinching force. This solution clearly shows the
role of the magnetic field in collimating the jet
but also the contribution of the charge separation to the
electric field and then to the decollimation.

The morphology of EMR jets is more affected by relativistic effects
than the one of IMR jets. This can be seen by 
comparing Fig. \ref{Jet_1_FigMorp01}b and Fig. 
\ref{Jet_1_Fig_Compare_morpholgie_classique_Relativiste}b.
In particular, the jet radius, or equivalently the expansion factor,  
asymptotically becomes more important in relativistic jets because of
the increase of the centrifugal and electric forces 
(Figs. \ref{Jet_1_FigForce_transverse_Rel}b and
\ref{Jet_1_FigCompare_Force__Classique}b).  Simultaneously, the
magnitude  of the Lorentz force decreases and
the thermal acceleration increases.

We can give a simple explanation to this relativistic effect. An
increase of gravity at the base of the jet induces a decrease of
the normalized electric current in the jet because $\bar{I}
\propto h_{ \star}$ where $\bar{I} = I/(c/2 r_{\star} B_{\star})$.
We have plotted the dimensionless electric current in Fig.
(\ref{Jet_1_Current_electrique_compare}) for the two EMR
solutions. The electric current $\bar{I}$ flows through a given
cross sectional area $S = \pi \varpi^2$ of the classical and
relativistic solutions.  We use the normalized electric current
because of the different scaling of the classical and the
relativistic solutions. The decrease of the current goes with a
decrease of the toroidal magnetic field and with an increase of
the expansion of the jet as explained in Sec. \ref{lambdaeffect}
and, consequently, with an increase of the poloidal velocity.
Note that this increase of the velocity corresponds to the
increase of the relativistic gravity as we already discussed. The new point 
is that, conversely
to hydrodynamical models, it also decollimates the flow.
On the other hand,  the magnetocentrifugal driving of the Poynting flux 
becomes weaker in
relativistic thermally driven winds, as expected.
Thus, as the rest mass increases, the
Poynting flux is getting weaker relatively to the other energies, ${\cal
E}_{\rm Poynt.} \ll m c^2$, while, the thermal
energy becomes relativistic and comparable to the rest mass $\left(w - m
c^2\right) \approx m c^2$.

\begin{figure}[h]
{\rotatebox{0}{\resizebox{8.0cm}{4.5cm}{\includegraphics{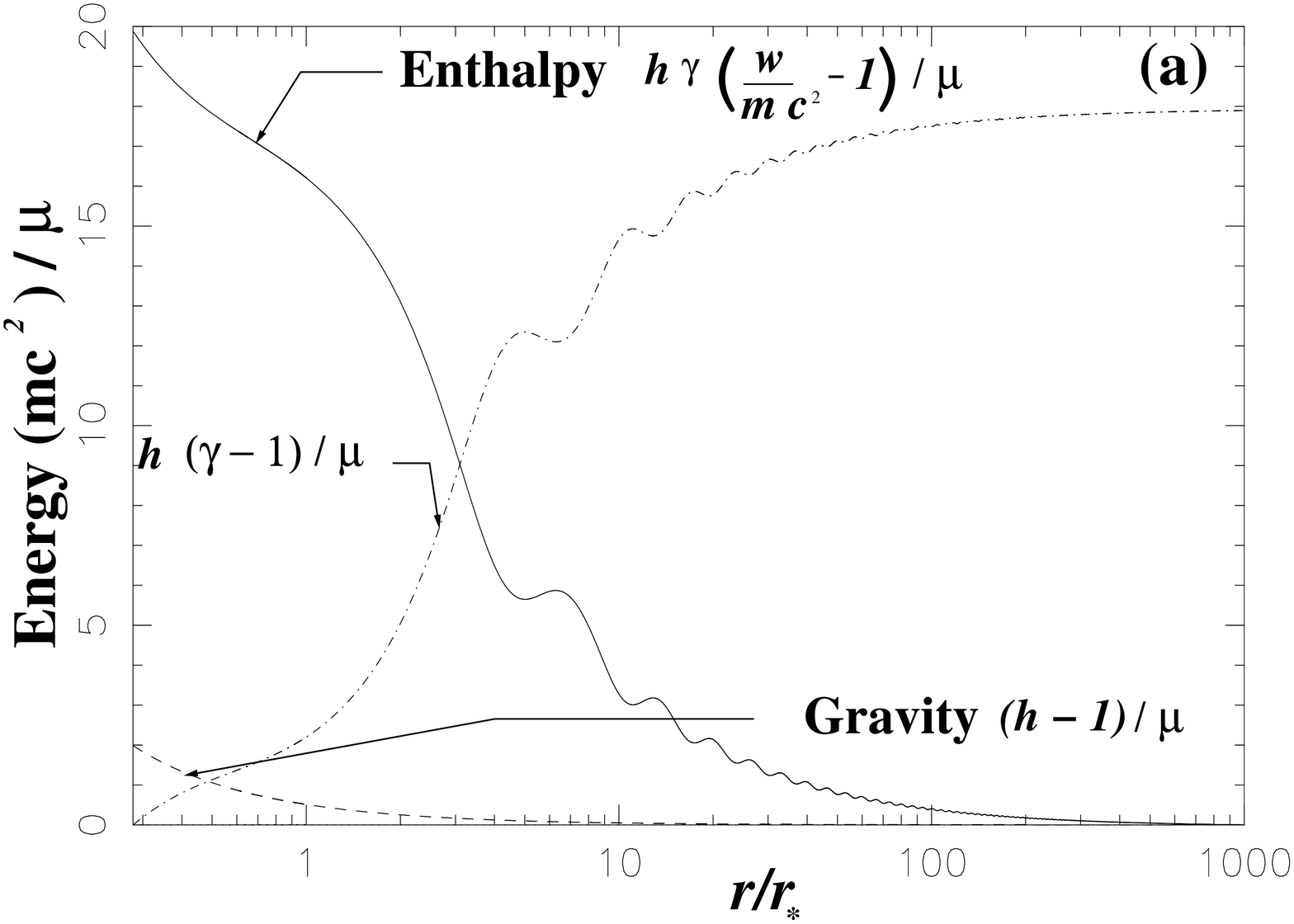}}}}
\rotatebox{0}{\resizebox{7.8cm}{4.5cm}{\includegraphics{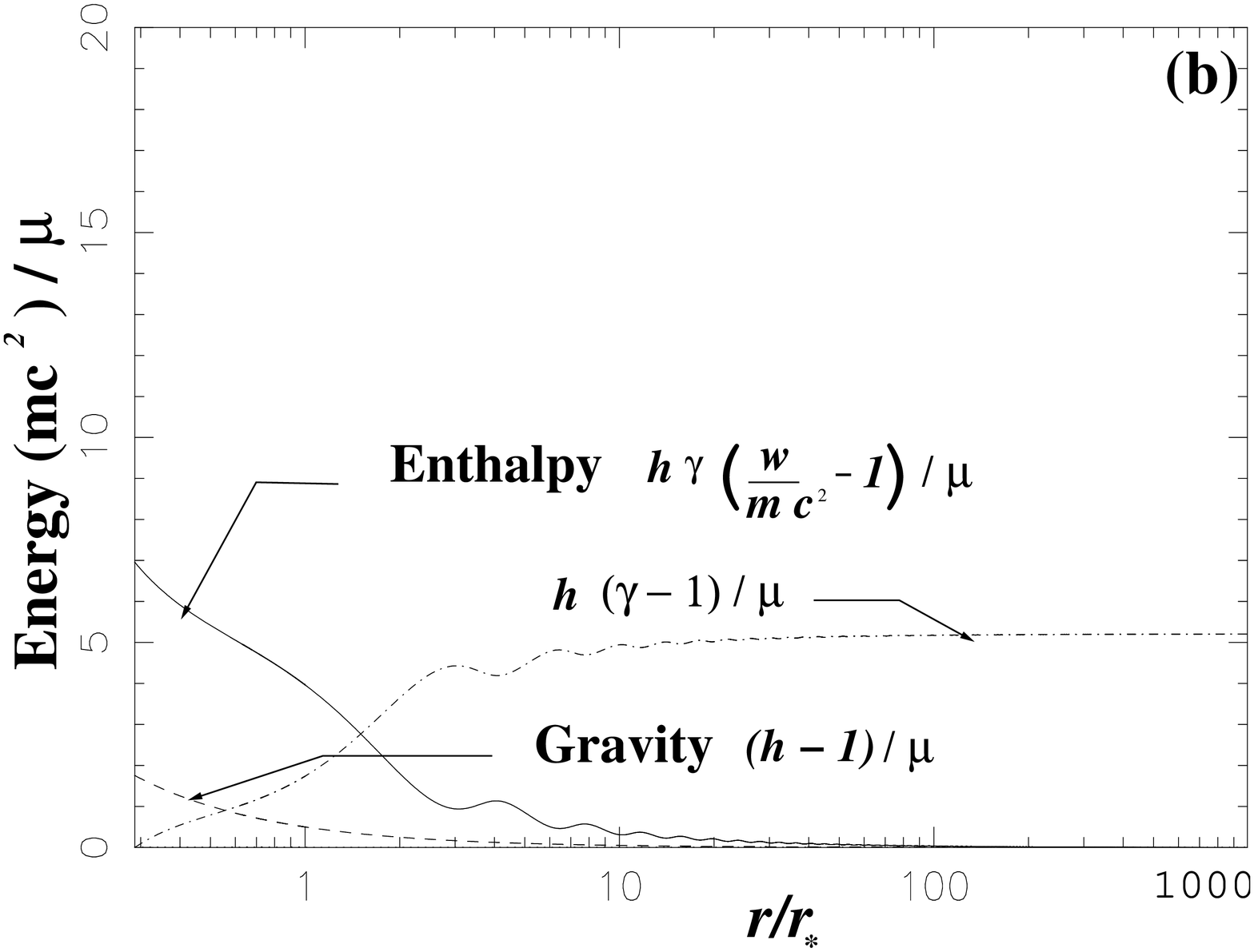}}}
\rotatebox{0}{\resizebox{8.0cm}{4.5cm}{\includegraphics{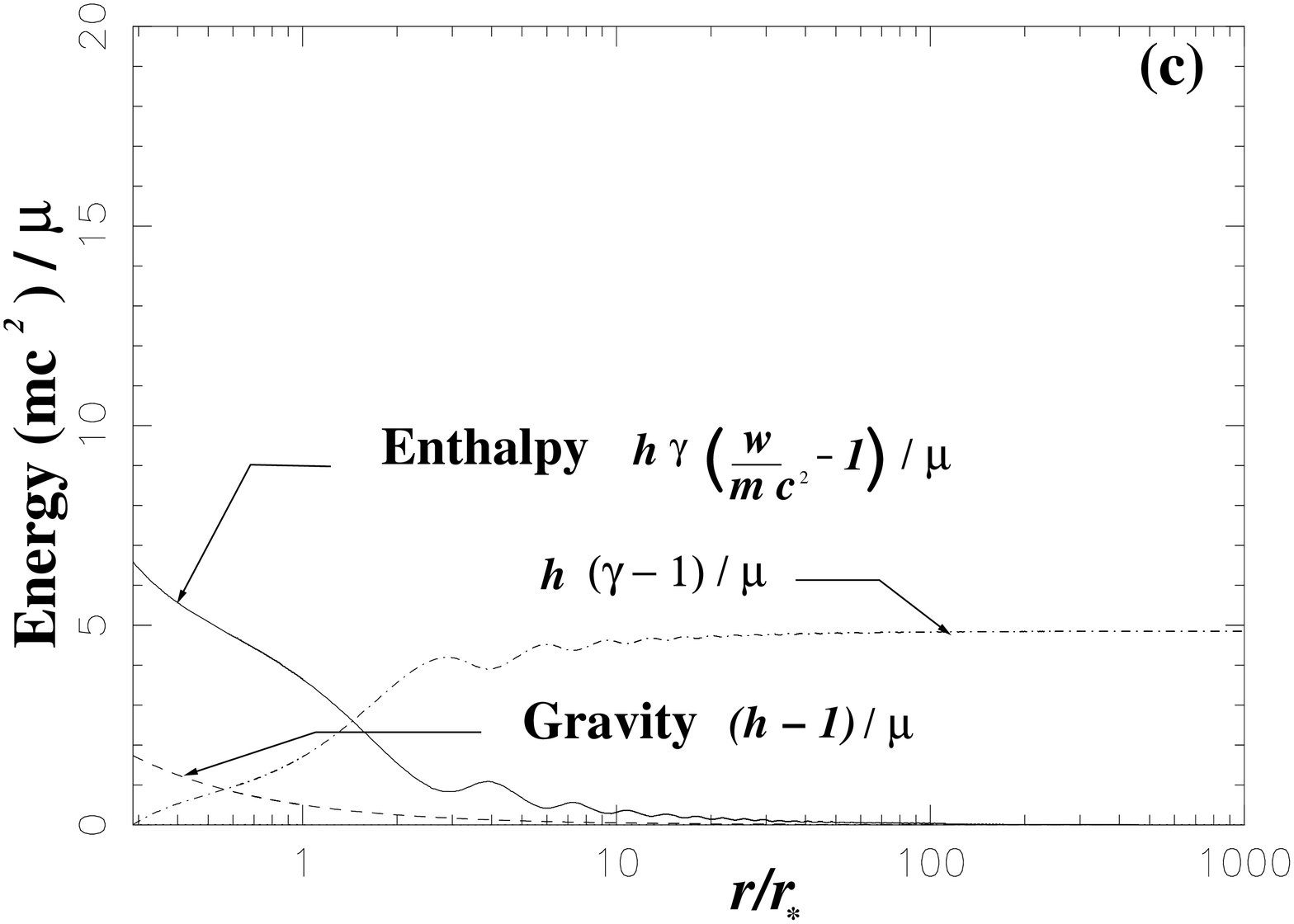}}}
\rotatebox{0}{\resizebox{8.1cm}{4.5cm}{\includegraphics{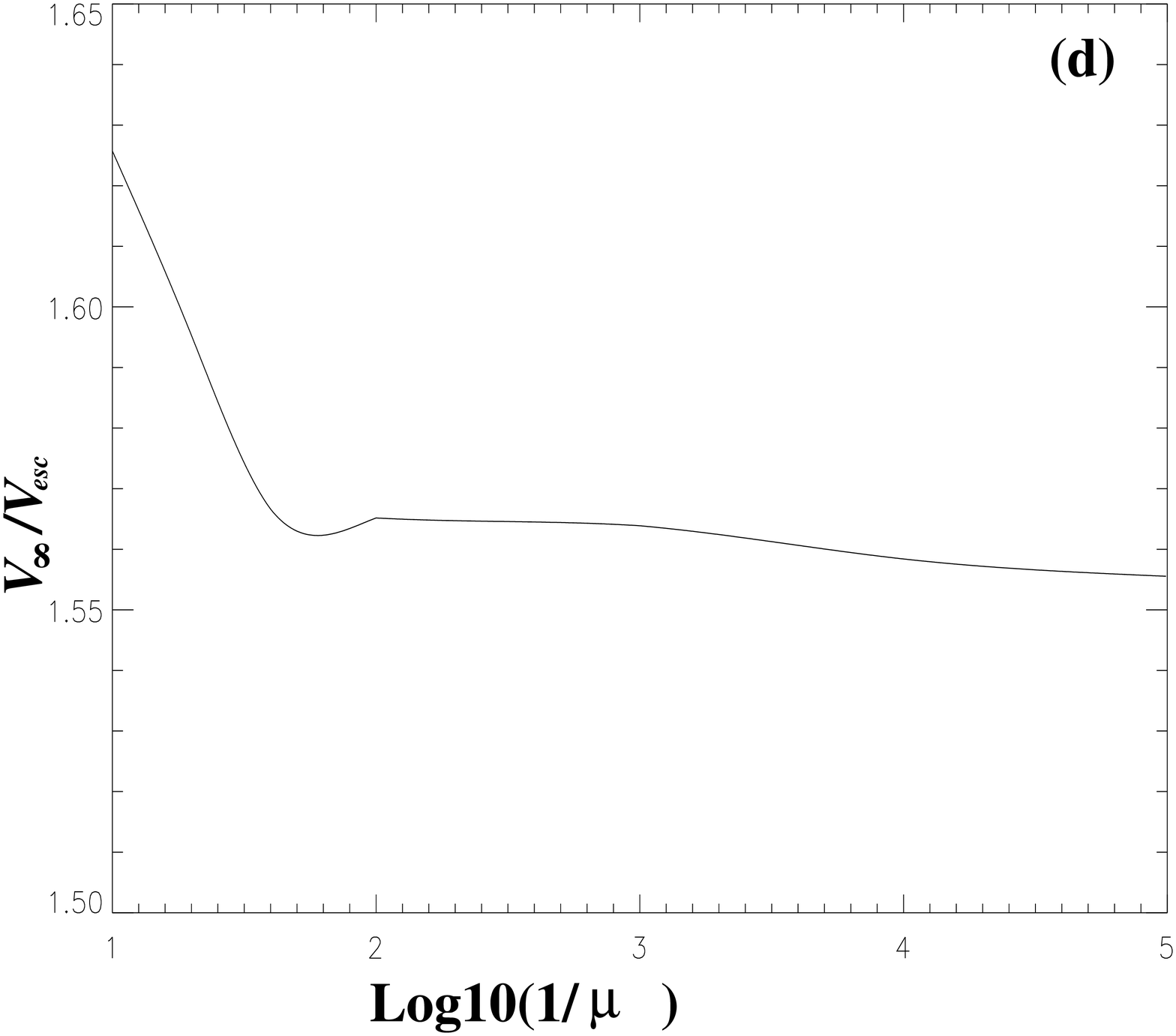}}}
\caption{ Plot of the energies normalized by the parameter $\mu$.
In (a) is shown the relativistic solution for an EMR, previously
displayed in Fig. \ref{Jet_1_Figenergy01}b with $\mu = 10^{-1}$.
In (c) is shown the corresponding non relativistic solution with
$\mu = 10^{-5}$. In (b) we plot an intermediate solution
 with $\mu = 10^{-2}$.  Plots (b) and (c) are identical which shows that for such
small values of $\mu$, the energies vary linearly with gravity, while this is not true in (a).
In (d) we plot the ratio between the  asymptotic velocity
 and the escape  velocity at the surface of the corona as a function of
 $\log{(1/\mu)}$.}
\label{Jet_1_Application_Gravite_effecte}
\end{figure}

The relativistic effects on the jet acceleration  become  remarkable only
for a distance between the Alfv\'en surface and the Schwarzschild surface
smaller than $100$ (ie. $\mu>0.01$, 
see Fig. \ref{Jet_1_Application_Gravite_effecte}a). For $\mu<0.01$,  
using this model, we see that outflows from a
star with  mass $1 M_{\odot}$ and  starting at $100 r_{G}$
are simply scaling with $\mu$ (Figs.
\ref{Jet_1_Application_Gravite_effecte}b and 
\ref{Jet_1_Application_Gravite_effecte}c). Similarly the ratio between the
energetics of the two jets are simply proportional to $\mu$. The scaling
just reflects the linear growth of the flow formation region with
gravity, i.e. with $\mu$.

Conversely, for $\mu>0.01$, the jet is formed at a distance smaller than
$100 r_{G}$, this linear scaling with $\mu$ of the dynamics and the energetics
does not hold any longer because of non linear relativistic effects.
The thermal energy converts more efficiently into kinetic energy
(Fig. \ref{Jet_1_Application_Gravite_effecte}a) as in the spherical case
(\cite{Melianietal04}). It increases even more because of the stronger 
expansion of the
relativistic jet in the super-Alfv\'enic region. In fact, in
the relativistic solution displayed in Fig. \ref{Jet_1_Application_Gravite_effecte}a,
collimation starts at $50$ Alfv\'en radii, while, in
the non relativistic solution,  Fig. \ref{Jet_1_Application_Gravite_effecte}c,
collimation occurs only at $10$ Alfv\'en radii.

In summary let us just point out that the mass of the
central object and the properties of the jet (acceleration, morphology and
energy) are not simply proportional to each other in the context of strong
gravitational fields.

\subsection{The effect of charge separation}

As we have seen, solutions obtained with this model are essentially thermally
driven winds but collimation is either thermal (pressure confinement)
or magnetic (toroidal magnetic pinching).
However, conversely to the non relativistic case an extra decollimating force
exists which is the electric force, despite that we neglect the
light cylinder effects.

The strength of the electric field results from the induction term
$V_{\rm p}B_\varphi /c$ and it is higher for higher magnetic fields. As
a matter of fact it gets more important when magnetic effects are important
and when the light cylinder is closer to the streamlines.
In relativistic flows where the poloidal velocity is of the order
of the light velocity, the contribution in the transverse direction
of this force increases in the super-Alfv\'enic domain.

Therefore, the contribution of this force, in relativistic jets
from EMRs, is of the order of the pinching force and the
centrifugal force. Conversely, in the non relativistic limit, the
poloidal velocity remains largely subrelativistic, $V_{\rm p}\ll
c$, and the electric field remains weak ($E \sim {V^2_{\rm p}
B^2_{\varphi}/c^2}/{\varpi}\rightarrow 0$;  see for
comparison Figs. \ref{Jet_1_FigForce_transverse_Rel}b and
\ref{Jet_1_FigCompare_Force__Classique}b). The electric field does not
affect the collimation of the non relativistic jet and the charge
separation can be neglected.

Conversely, the effects of the electric force are negligible in
pressure confined jets from IMRs as the magnetic effects
themselves are very weak or completely  negligible,
${B_{\varphi}^2}/{\varpi} \ll {P}/{\varpi}\Rightarrow \rho E
\ll{P}/{\varpi}$.

Parallely, the effects of the electric force on the
acceleration of the jet are very weak even for EMRs.  The
correction introduced on the asymptotic speed is negligible.
This is because the electric force is perpendicular to the streamlines
and, so, it affects mainly the morphology of the jet.

\section{Conclusion}

In this paper, we have investigated the problem of the formation
and collimation of relativistic jets. We have explored these
problems by means of a semi-analytical model, which is the first
meridional self-similar model of relativistic jets, including
general relativistic effects. We constructed it on the basis of
the classical model developed in ST94 to study jets from young
stars. We have made an extension of this model for a black hole
characterized by weak angular momentum, $a = {J}{{\cal G}/( {\cal
M}_{\bullet}^2/c}) \ll 1$. Therefore, we treated the problem of
GRMHD outflows in a Schwarzschild metric. Moreover, we
concentrated our efforts on modelling the jet close to its polar
axis. In the construction of this model, we were limited to
describe jets possessing a weak rotation velocity compared to the
speed of light  and we neglected the effects of the light cylinder.
Thus, we restricted our study to thermally  accelerated jets with
a weak contribution of the Poynting flux.
However, in the collimation of the jet, both electric and magnetic
terms are comparable to the inertial and pressure gradient ones.
We have also studied  the collimation effects by magnetic and
thermal forces and the decollimation effect of centrifugal and
electric forces.
 Our model is restricted to outflow solutions only, and we do not
address to the problem of the origin of the hot coronal plasma.

We found that the influence of the electric force and the charge
separation in the
jet depends on the collimation regime. In the case of EMRs
where jets are magnetically collimated, the electric decollimating force
plays an important role. This force is of the order of the magnitude
of the centrifugal force and of the magnetic pinching.
Therefore, relativistic jets from EMRs are less collimated than
their non relativistic counterparts.
Conversely, jets from IMRs,
where collimation is mainly of thermal origin, are not very sensitive
to the electric decollimation.
 In this type of jets, the contribution of $\rho_e  \vec{E}$ is balanced
by the increase of the external pressure.

We have also used the model to compare classical jets to
relativistic jets. We undertook this comparative study by changing
the free parameter $\mu$. In fact, this new parameter gives in the
relativistic model the space curvature which is induced by gravity
near the central object. We used typical values of $\mu$ from
$\mu\sim 10^{-5}$ for Jets from Young Stellar Objects to $\mu \sim
0.1$ for jets from compact sources. We found that the difference
between these two types of jets is only a scaling effect on $\mu$
for $\mu < 10^{-2}$. In this case, the spatial dimensions and
energies are linear functions of $\mu$. For $\mu> 10^{-2}$, the
relativistic effects increase together with the thermal
acceleration of the jet. Simultaneously, strongly relativistic
effects tend to decollimate the jets because of the decrease of
the electric current density.

To conclude we have seen that relativistic effects and
particularly relativistic gravity tend to enhance the thermal
acceleration (as in Meliani et al. 2004)  and reduce the magnetic
collimation of the jet (as in Bogovalov \& Tsinganos 2005). Still
collimation can be obtained either by thermal or magnetic means
but relativistic effects lower the efficiency of the magnetic
rotator. This means that despite  quantitative changes, we can use
this generalized model to verify, for the classification of AGN
jets, the conjectures given in Sauty et al. (2001) using the non
relativistic model. Mainly we see that the observed types of jets
from radio loud galaxies can be connected to the efficiency of
their magnetic rotator and to the environment of the host galaxy.
By using the present model, a more precise and quantitative
analysis of the observed jets will be presented in a following
paper.

\acknowledgements{ We acknowledge financial support for our visits
from the French Foreign Office and the Greek General Secretariat of 
Research and Technology (Program PLATON), from the European RTN
JETSET (MRTN-CT-2004-005592) and the Observatoire de Paris. 
E.T. acknowledges financial support by the Italian
Ministry for Education, University and
Research (MIUR) under the grant Cofin 2003/027534-002. 
Finally, part of this work has been also supported by the European RTN
ENIGMA (HPRN-CY-2002-00231).} 

\appendix

\section{Ordinary differential equations}\label{appendixA}

\begin{eqnarray}
\label{dPdR}
\frac{{\rm d}  \Pi}{{\rm d} R} &=&
-\frac{2}{{h_{} }^2}\frac{1}{G^4}
\left(\frac{{\rm d}  M^2}{{\rm d} R}+\frac{F-2}{R}
 M^2\right)
\nonumber\\&&
-\frac{1}{h_{}^4 R^2 M^2}\left(\nu^2 h_{\star}^4-\mu\frac{M^4}{G^4}\right)
\,,\end{eqnarray}

\begin{equation}
\label{A2}
\frac{{\rm d}  M^2}{{\rm d} R} = \frac{{\cal N}_M}{{\cal D}}
\label{dMdR}
\,,\end{equation}

\begin{equation}
\frac{{\rm d}  F}{{\rm d} R} = \frac{{\cal N}_F}{{\cal D}}
\label{dFdR}
\,,\end{equation}

\begin{equation}
{\cal D} = -\left(1+\kappa\frac{R^2}{G^2}\right) D
+{\lambda^2R^2}\frac{N_B^2}{D^2}+\frac{h_{}^4 F^2}{4h_{\star}^2}
\,,\end{equation}

\begin{eqnarray}\label{Jet_1_Int_dM2dR}
{\cal N}_M &=& \frac{1}{4h_{\star}^2}\frac{{M}^4}{R}
\left(8+8\kappa\frac{R^2}{G^2} - 2F -4F\kappa\frac{R^2}{G^2}
- h_{}^2F^2\right)
\nonumber\\&&
+\frac{{h_{}^2 }}{h_{\star}^2}\frac{M^2}{R}
\left[-2-2\kappa\frac{R^2}{G^2}
{ -F\frac{\lambda^2\mu}{\nu^2}}\frac{h_{\star}^2}
{h_{}^2}R^2+F\kappa\frac{R^2}{G^2}
\right.\nonumber\\&&\left.
+\frac{h_{}^2}{4} \left ( 1+\frac{\mu}{R h_{}^2} \right) F^2 
+\frac{h_{}^2}{8}F^3\right]
\nonumber\\&&
-D\frac{\nu^2 h_{\star}^4}{2{h_{} }^2}\frac{G^2}{ M^2}(\delta-\kappa)
+\frac{h_{}^2}{h_{\star}^2}\frac{1}{2}\kappa \Pi R G^2F M^2
\nonumber\\&&
-\frac{M^2}{2h_{}^2}\frac{\mu}{R^2}\left(1+
\kappa\frac{R^2}{G^2}\right)D
+\lambda^2\frac{N_B N_V}{D^2}\mu
\nonumber\\&&
+\lambda^2R\left(\frac{N_B^2}{D^2}-\frac{h_{}^2}{2 M^2}
\frac{N_V^2}{D^2} \right)
\left[2M^2+h_{}^2 (F-2)\right]
\nonumber\\&&
-\lambda^2 R\, h_{}^2 (F-2)\frac{N_B}{D}
+\mu \lambda^2 \frac{G^2 h_{\star}^4 N_B}{h_{}^2 M^2}
\,,
\end{eqnarray}

\begin{eqnarray}
{\cal N}_F&=&
-\frac{ M^2F}{R h_{\star}^2}\left[\left(1-\frac{F}{2}\right)
\left(1+\kappa\frac{R^2}{G^2}+
{\lambda^2R^2}\frac{N_B^2}{D^3}\right)
\right.\nonumber\\&&\left.
+\frac{F}{2}\left(1-h^2_0\frac{F}{2}\right)\right]
\nonumber\\&&
+\frac{1}{R}\left(\frac{h_{} }{h_{\star}}\right)^2
\left(1+\kappa\frac{R^2}{G^2}-{\lambda^2 R^2}
\frac{N_B^2}{D^3}-\frac{h_{}^2}{4}F^2\right)
\nonumber\\&&
\left(F^2-F-\frac{2}{h_{}^2}
{- \frac{{h_{\star}}^2}{{h_{}}^4}\frac{\lambda^2\mu}{\nu^2}4R^2}\right)
\nonumber\\&&
+\left(\frac{2R\Pi G^2\kappa}{h_{\star}^2}+\frac{\mu F}{h_{\star}^2 R^2} \right)
\nonumber\\&&
\left(1+\kappa\frac{R^2}{G^2}- {\lambda^2 R^2}\frac{N_B^2}{D^3}-\frac{h_{}^2}{4}F^2\right)
\nonumber\\&&
-\frac{\nu^2 G^2 h_{\star}^2}{2 M^2 h_{}^2}F(\delta-\kappa)
-\frac{\mu F M^2}{2 R^2h_{}^2 h_{\star}^2}
\left(1+\kappa\frac{R^2}{G^2}\right)
\nonumber\\&&
+\frac{\lambda^2}{h_{\star}^2}\left(\frac{N_B^2}{D^2}
-\frac{N_B}{D}\right)\frac{F}{D}\mu
-\frac{\lambda^2 R h_{}^2}{h_{\star}^2}F(F-2)\frac{N_B}{D^2}
\nonumber\\&&
+\frac{4\lambda^2R}{h_{}^2}\left(\frac{N_B^2}{D^2}-\frac{1}{2 M^2}
h_{}^2\frac{N_V^2}{D^2}\right)
\nonumber\\&&
\left(1+\kappa\frac{R^2}{G^2}-\lambda^2 R^2\frac{N_B^2}{D^3}
-\frac{h_{}^2}{2}F\right)
\nonumber\\&&
+\mu \lambda^2 \frac{G^2 h_{\star}^2 F }{h_{}^2 M^2} \frac{N_B}{D}
\label{Jet_1_Int_dFdR}
\,.
\end{eqnarray}

\onecolumn

\section{The forces on the plasma}\label{AppendixB}
The momentum equation can be written as
\begin{equation}
-\frac{n}{\Psi_A^2 w/c^2} \left(\vec{U}\cdot \vec{\nabla} \right) \vec{U}
-\vec{\nabla} P + \frac{\left(\vec{\nabla}\cdot\vec{E}\right)}{4 \pi} \vec{E}
+ \frac{\left[\vec{\nabla}\times\left( h_{} \vec{B} \right)\right]}{4 \pi h_{}} \times \vec{B}
-\gamma^2 n w \vec{\nabla} \ln h_{} =0
\,,\end{equation}
where the generalized velocity $\vec{U}$ is given by
\begin{equation}
\vec{U}_p = \Psi_A \gamma \frac{w}{c^2}\vec{V}_p =
\frac{M^2}{h_{}}\vec{B}_p \,,
\end{equation}
\begin{equation}
U_\varphi = \Psi_A \gamma \frac{w}{c^2} V_\varphi
= \lambda h_{\star} B_{\star} \frac{R}{G^2}
\left[\frac{M_{}^2-h_{\star}^2 G^2
(1-x_{\rm A}^2)}{M_{}^2-h_{}^2 + G^2 h_{\star}^2 x_{\rm A}^2}\right]
\sin\theta \,.
\end{equation}
In the following we give the expressions of the various terms.


\subsection{Advection force}
\begin{eqnarray}
-\frac{n}{\Psi_A^2 w/c^2}
\left[\left(\vec{U}\cdot\vec{\nabla}\right) \vec{U} \right] \cdot \vec{e}_{R} &=&
-\frac{B_{\star}^2}{4\pi r_\star G^4}
\left\{ \frac{1}{h_{}^2} \frac{{\rm d} M^2}{{\rm d} R} +
\frac{M^2}{h_{} R} \left(F - 2 - \frac{\mu}{2 R h_{}^2}\right)
\right . \nonumber\\
&& \left .  + \sin^2\theta \left[ - \frac{1}{h_{}}  \frac{\rm d M^2}{\rm d R}
- \frac{M^2}{h_{} R} \left(h_{}^2 \frac{F^2}{4} + \frac{F}{2} - 2
+ \frac{\mu}{2 R h_{}^2}\right) - \lambda^2 h_{} h_{\star}^2 \frac{R}{M^2}
 \frac{N_{V}^2}{D^2}\right]\right\}\,,
\label{Adv_Force_R}
\end{eqnarray}

\begin{eqnarray}
-\frac{n}{\Psi_A^2 w/c^2}
\left[\left(\vec{U}\cdot\vec{\nabla}\right) \vec{U} \right] \cdot \vec{e}_{\theta} &=&
\frac{B_{\star}^2}{4\pi r_\star G^4} \sin\theta \cos\theta \nonumber\\
&& \left[ \frac{F}{2} \frac{{\rm d} M^2}{{\rm d} R} + \frac{M^2}{R} \frac{F}{2}
\left(\frac{F}{2}  - 1\right) + \frac{M^2}{2} \frac{{\rm d} F}{{\rm d} R}
+ \frac{\lambda^2 h_{\star}^2 R}{M^2} \frac{N_{V}^2}{D^2}\right]
\end{eqnarray}

\subsection{Pressure force}
\begin{equation}
\label{Annex_Force_Pressionr}
\vec{f}_{Pr}^{R}=-\frac{h_{}}{r_\star}
\frac{\partial P}{\partial R}=-
\frac{h_{}}{8} \frac{B_{\star}^2}{\pi r_\star G^4}
\left[\frac{{\rm d} \Pi}{{\rm d} R} G^4
\left(1+\kappa \frac{R^2}{G^2}\sin^2{\theta}\right)+\Pi {F}
\kappa R G^2 \sin^2 \theta \right] \vec{e}_{R}\,,
\end{equation}
\begin{equation}
\label{Annex_Force_Pressiontheta}
\vec{f}_{Pr}^{\theta}=-\frac{1}{r_\star R}\frac{\partial P}
{\partial \theta}=-\frac{B_{\star}^2}{4\pi r_\star G^4}RG^2
\Pi \kappa \sin{\theta} \cos{\theta} \vec{e}_{\theta}\,.
\end{equation}

\subsection{Electric force}
\begin{equation}
\vec{f}_{E}^{R}=\frac{B_\star^2}{4\pi r_\star G^4} \frac{h^2_{0\star}}{h_{}} F
\lambda^2 \frac{V^2_{\star}}{c^2} R \sin^2\theta \vec{e}_{R}
\,, \end{equation}
\begin{equation}
\vec{f}_{E}^{\theta}=  \frac{B_{\star}^2}{2\pi r_\star G^4}
\left(\frac{h_{\star}}{h_{}}\right)^2\lambda^2
\frac{V^2_{\star}}{c^2}R\sin\theta\cos\theta \vec{e}_{\theta}
\,, \end{equation}

\subsection{Magnetic force}
\subsubsection{Magnetic hoop stress}
\begin{eqnarray}\label{Annex_Force_Pincement_Bphi}
\vec{f}_{S, B\varphi}^{R}=
- \frac{1}{4\pi r_\star}B_{\varphi}^2
\left(\frac{h_{}}{R}+\frac{dh_{}}{dR}\right) \vec{e}_{R} &=&
-\frac{B_{\star}^2}{4 \pi r_\star G^4} \left(\frac{h_{}}
{R}+\frac{dh_{}}{dR}\right)
\frac{\lambda^2h_{\star}^2}{h_{}^2}
{\left[{\frac{{(\frac{h_{}}{h_{\star}})}^2 -G^2}
{\frac{M^2}{h_{ \star}^2} -{(\frac{h_{}}{h_{\star}})}^2}}\right]}^2 R^2
\sin^2{\theta} \vec{e}_{R} \,
\end{eqnarray}
\begin{equation}
\vec{f}_{S, B\varphi}^{\theta} = -\frac{1}{4 \pi r_\star R} B_{\varphi}^2
\cot{\theta} \vec{e}_{\theta}
 = -\frac{\lambda^2 B_{\star}^2}{4 \pi r_\star G^4 }
\frac{N_B^2}{D^2} \frac{{h_{ \star}}^2}{{h_{}}^2}  R \sin{\theta}
\cos{\theta}\vec{e}_{\theta}\, .
\end{equation}
\begin{eqnarray}\label{Annex_Force_Pincement_BP}
\vec{f}_{S, Bp}^{R} &=& \left[
-\frac{1}{4 \pi r_\star} B_{\theta}^2 \left(\frac{h_{}}{R}
+ \frac{{\rm d}h_{}}{{\rm d}R}\right)  +
\frac{B_{\theta}}{4 \pi r_\star R}\frac{\partial B_r}{\partial \theta} \right] \vec{e}_{R}\nonumber\\
&=& \left[
-\frac{B_{\star}^2}{16 \pi r_\star G^4}  h_{}^2 F^2
\left(\frac{h_{}}{R} + \frac{{\rm d}h_{}}{{\rm d}R}\right)
\sin^2{\theta}
+  h_{} \frac{B_{\star}^2}{8 \pi r_\star G^4} \frac{F}{R} \sin^2{\theta} \right]
\vec{e}_{R}\,,\\
\vec{f}_{S, Bp}^{\theta} &=& \left[\frac{1}{4 \pi r_\star}B_{r} B_{ \theta}
\left( \frac{{\rm d}h_{}}{{\rm d}R}  + \frac{h_{}}{R}\right)
+\frac{h_{} B_r}{4 \pi r_\star} \frac{\partial B_\theta}{\partial R}\right] \vec{e}_{\theta}\nonumber\\
 &=& -\frac{B_{\star}^2}{8 \pi r_\star} \frac{h}{G^4} 
\left[h\frac{{\rm d}F}{{\rm d} R} + h\frac{F^2}{R} - h \frac{F}{R} + F 
\frac{\mu}{h R^2}\right] \sin\theta \cos\theta \vec{e}_{\theta} \,.
\end{eqnarray}

\subsubsection{Magnetic gradient pressure}

\begin{eqnarray}\label{Annex_Force_pressionB_toroïdale}
\vec{f}_{Pr, B\varphi}^{R}& = & -\frac{h_{}}{8 \pi r_\star}
\frac{\partial B_{\varphi}^2}{\partial R}
\vec{e}_{R} =
-\frac{h_{ \star}^2}{h_{}^2}\frac{\lambda^2 B_{\star}^2}{ 4 \pi r_\star G^4}
 \left[\frac{\left(\frac{{h_{}}}{h_{ \star}}\right)^2 - G^2}
{{(\frac{{h_{}}}{h_{ \star}})}^2 - \frac{M^2}{h_{, \star}}}\right]^2
R^2\sin^2{\theta}\nonumber\\
&&\left[\frac{{h_{}}}{R}-\frac{\mu}{2 R^2 h_{}}
+ h_{}\frac{F - 2}{R} +
h_{}\frac{\frac{\mu}{R^2 h_{ \star}^2}+\frac{G^2}{R}(F-2)}
{\left(\frac{{h_{}}}{h_{ \star}}\right)^2 - G^2}
- h_{}
\frac{\frac{\mu}{R^2 h_{\star}^2}-\frac{{\rm d}M^2}
{h_{\star}^2 {\rm d}R}}{\left(\frac{h_{}}{h_{\star}}\right)^2
-\frac{M^2}{h_{ \star}^2}}\right] \vec{e}_{R} \,,\\
\vec{f}_{Pr, B\varphi}^{\theta}& = & -\frac{1}{8 \pi r_\star R}
\frac{\partial B_{\varphi}^2}{\partial \theta} \vec{e}_{\theta} = -\lambda^2
\frac{B_{\star}^2}{4 \pi r_\star G^4 } \frac{N_B^2}{D^2}
\left(\frac{h_{ \star}}{h_{}}\right)^2 R \sin{\theta} \cos{\theta} \vec{e}_{\theta} \,  .
\end{eqnarray}

\begin{eqnarray}\label{Annex_Force_pressionB_pololïdale}
\vec{f}_{Pr, Bp}^{R}& = & -\frac{h_{}}{8 \pi r_\star}
\frac{\partial B_{\theta}^2}{\partial R} \vec{e}_{R}
\nonumber\\
&=& -h_{}^2 \frac{B_{\star}^2}{16 \pi r_\star G^4} F^2
\left[\frac{\mu}{2R^2 h_{}} + h_{}
\left(\frac{1}{F}\frac{{\rm d}F}{{\rm d}R}
+\frac{F-2}{R}\right)\right] \sin^2\theta \vec{e}_{R} \,,
\\
\vec{f}_{Pr, Bp}^{\theta}& = & -\frac{1}{8 \pi r_\star R}
\frac{\partial B_r^2}{\partial \theta} \vec{e}_{\theta}
 = \frac{B_{\star}^2}{4 \pi r_\star G^4} \frac{1}{R}
\sin{\theta} \cos{\theta} \vec{e}_{\theta} \,.
\end{eqnarray}

\subsection{Gravity force}
\begin{equation}
-\gamma^2 n w \vec{\nabla} \ln h_{}=
-\frac{\gamma^2 n w}{r_\star} \frac{{\rm d} h_{}}{{\rm d} R} \vec{e}_{R}
= -\frac{B_{\star}^2}{8\pi r_\star} \frac{h_{\star}^4}{h_{}^3}
\frac{\nu^2}{ R^2 M^2}
\left(1 + \delta \alpha - 2 \frac{\mu \lambda^2}{\nu^2}
\frac{N_B}{D} \alpha \right) \vec{e}_{R}\,,
\label{Gra_Force_R}
\end{equation}
\twocolumn

\end{document}